\documentclass{article}
\usepackage{emulateapj}
\usepackage{float,epsfig}
\usepackage{natbib}
\newcommand{\LA}{\mbox{\raisebox{-0.6ex}{$\stackrel{\textstyle<}{\sim}$}}}
\newcommand{\msun}{M$_{\odot}$}
\newcommand{\msfr}{M$_{\odot}$~yr$^{-1}$}
\newcommand{\ergl}{ergs~s$^{-1}$}

\newcommand{\DTF}{{$D_{25}$}}

\newcommand{\ha}{H$\alpha$}
\newcommand{\hi}{H{\sc i}}

\newcommand{\cxo}{{\sl Chandra}}
\newcommand{\xmm}{{\sl XMM-Newton}}
\newcommand{\hst}{{\sl Hubble}}
\newcommand{\ros}{{\sl ROSAT}}
\newcommand{\etal}{et al.}
\newcommand{\elip}{{$D_{25}$}}

\slugcomment{Submitted to Astrophysical Journal}

\begin{document}

\title{A Complete Sample of ULX Host Galaxies}

\author{
Douglas~A.~Swartz\altaffilmark{1}
Roberto~Soria\altaffilmark{2}, 
Allyn~F.~Tennant\altaffilmark{3}, and
Mihoko~Yukita\altaffilmark{4} }

\altaffiltext{1}{Universities Space Research Association,
    NASA Marshall Space Flight Center, VP62, Huntsville, AL, USA}
\altaffiltext{2}{Curtin Institute of Radio Astronomy,
    Curtin University, 1 Turner Avenue, Bentley WA 6102,  Australia}
\altaffiltext{3}{Space Science Office,
    NASA Marshall Space Flight Center, VP62, Huntsville, AL, USA}
\altaffiltext{4}{Department of Physics \& Astronomy,
    University of Alabama, Tuscaloosa, AL, USA}

\begin{abstract}

One hundred seven ultraluminous X-ray (ULX) sources with 0.3$-$10.0~keV 
luminosities in excess of $10^{39}$~\ergl\ are identified in a complete 
sample of 127 nearby galaxies. The sample includes all galaxies within
14.5~Mpc above the completeness limits of both the Uppsala Galaxy Catalog 
and the {\sl Infrared Astronomical Satellite}\/ survey.
The galaxy sample spans all Hubble types, a
four decade range in mass, $7.5<\log(M/{\rm M}_{\odot})<11.4$, and in
star-formation rate, $0.0002<{\rm SFR (M}_{\odot}/{\rm yr)}\le 3.6$.
ULXs are detected in this sample at rates of one per $3.2\times10^{10}$~\msun,
 one per $\sim$0.5 \msfr\ star-formation rate, and one per 57~Mpc$^3$
 corresponding to a luminosity density of 
 $\sim$2$\times$10$^{37}$ erg~s$^{-1}$~Mpc$^{-3}$. 
At these rates we estimate as many as 19 additional ULXs remain undetected 
 in fainter dwarf galaxies within the survey volume. 
An estimated 14 or 13\%, of the 107 ULX candidates 
 are expected to be background sources.
The differential ULX luminosity function 
 shows a power law slope $\alpha \sim -1.2$ to $-$2.0
 with an exponential cutoff at $\sim$20$\times$10$^{39}$~\ergl\ 
 with precise values depending on the 
 model and on whether the ULX luminosities are estimated from their observed 
 numbers of counts or, for a subset of candidates, from their spectral shapes.
Extrapolating the observed luminosity function predicts at most one very luminous ULX, 
$L_{\rm X} \sim  10^{41}$~\ergl, 
within a distance as small as 100~Mpc.
The luminosity distribution of ULXs within the local universe
 cannot account for the recent claims of luminosities in excess of 
 2$\times$10$^{41}$~\ergl\ 
 requiring a new population class to explain these extreme objects.

\end{abstract}

\keywords{galaxies: general --- surveys --- X-rays: binaries ---
X-rays: galaxies --- X-rays: general}

\section{Introduction}

Ultraluminous X-ray sources (ULXs) 
 are defined as non-nuclear point-like objects in external galaxies,
with apparent isotropic X-ray luminosity higher than that of stellar-mass
Galactic black holes: typically, $L_{\rm X} > 10^{39}$~\ergl\
in the 0.3$-$10~keV band. By this empirical definition, ULXs
may encompass extreme examples of high-mass and low-mass X-ray
binaries, recent supernovae, rotation-powered young pulsars
and perhaps intermediate-mass black holes. In practice, supernovae
are excluded, and high-energy pulsars are not expected to be a significant
population at these luminosities (Perna \& Stella 2004; Perna \etal\ 2008). 
So, the study of ULXs concentrates
exclusively on accretion systems. Blazar-like beaming is ruled out
for at least some of the brightest sources,
hence the consensus is that most ULXs do have high luminosites
and high mass accretion rates. This has led to two diverging scenarios
with circumstantial observational support: 
(1) near-Eddington or moderately super-Eddington sources powered
by stellar black holes, $M \la 80$ \msun;
(2) sub-Eddington sources
powered by high mass accretors, with $M \ga 100$ \msun.
In scenario (1), the high accretion rate implies that
standard thin-disk models are not appropriate, and have to be
replaced by geometrically thicker disks (e.g., slim disks, Abramowicz 1988; Kawaguchi 2003,
 Poutanen \etal\ 2007; Gladstone \etal\ 2009)
with massive 
outflows.\footnote{Hence, the standard disk scaling relations between temperature, luminosity
and black hole mass, well-tested for (sub-Eddington) Galactic black holes,
would not be applicable to ULXs.}
This introduces the possibility of mild
geometrical beaming inside a conical outflow (King 2009 and references therein), which
would reduce the required black hole mass and the need to violate
the classical Eddington limit.
The accreting intermediate-mass black holes
of scenario (2) might come from Pop III stars (Madau \& Rees 2001), from the collapsed
core of young, massive star clusters or molecular cores (Portegeis~Zwart \etal\ 2004),
or from the nucleus of accreted satellite galaxies (King \& Dehnen 2005).
But the likelihood that such exotic objects can capture a donor
star and form an X-ray luminous system is low (Baumgardt \etal\ 2006; Madhusudhan \etal\ 2006).

Ultimately, the question we want to answer is: are ULXs an extreme
example of well-known X-ray binaries (the high-accretion-rate and
high-black-hole-mass tail of the X-ray binary distribution), or do they require
a new, physically distinct, population of accreting black holes not formed
from single-star processes in the current epoch?
We do not have direct estimates of black hole masses with which these
 two scenarios could be distinguished. Since X-ray luminosity is the defining
property of this class, and probably the least model dependent
quantity, the question we ask instead is whether there is a maximum ULX luminosity.
If ULXs are powered by stellar black holes, we expect their maximum luminosity to be 
at $L_{\rm X} \sim 10^{40}$~\ergl\ within a factor of a few corresponding to the 
 Eddington limit of the maximum stellar black hole mass. 
If ULXs are powered by sub-Eddington intermediate-mass black holes, we expect
luminosites of 
$\sim 10^{42}$ ergs s$^{-1}$ and above for black hole masses $\gtrsim 10^4 $ \msun.

Moreover, the same $L_{\rm X}$
requires different mass accretion rates in the two scenarios,
and hence we predict a different lifetime of the brightest sources.
For super-Eddington sources,
$L_{\rm X} \propto L_{\rm Edd} (1+ \frac{3}{5} \ln (\dot{M}/\dot{M}_{\rm Edd}))$; a luminosity
$L_{\rm X} \sim 10^{40}$ ergs s$^{-1}$ requires
$\dot{M} \sim 2 \times 10^{-5} $ \msun\ yr$^{-1}$ if coming from
a 25~\msun\ black hole. 
For sub-Eddington sources, $L_{\rm X} \propto 0.1 \dot{M} c^2$;
the same luminosity $L_{\rm X} \sim 10^{40}$ ergs s$^{-1}$
requires an accretion rate of only $\dot{M} \sim 2 \times 10^{-6}$~\msun\
yr$^{-1}$ for black hole masses $> 100$~\msun. 

Of course, the luminosity of X-ray binaries is rarely steady; one cannot equate 
a single observed luminosity of a particular object to a definite mass with any certainty.
But, if ULXs {\sl as a class} are powered by stellar-mass black holes, then
the high-end of their luminosity distribution would have a sharp downturn
because of the combined effect of upper black hole mass limit and shorter active lifetime.
Conversely, if ULXs are intermediate-mass black holes, then no luminosity cutoff would be expected.
 
The best approach to addressing these predictions
 is to study a large sample of ULXs in a variety of galaxies
in order to 
determine their luminosity distribution.
%
An ideal survey would
 include every galaxy within a fixed distance.
Once the ULX phenomenon are understood and quantified
within such a complete sample,
 the results can be extended to studies of more distant objects 
 where only the integrated X-ray light from entire galaxies 
 can be measured. 
This is the goal of the present study.
Unfortunately, there are no measurements that go deep enough
 to catalog all the low-mass dwarf galaxies out to a reasonable distance.
We can still achieve most of the scientific advantages of a complete survey
 by accounting for the missing population 
 by using well-defined and
 quantifiable selection criteria.
Thus, our aim is to produce a statistically rigorous sample of
 ULX candidates,
 to provide precise astrometric positions, and
 to estimate their X-ray luminosities
 using a carefully defined sample of nearby galaxies. 
 
The survey galaxy sample is defined in \S \ref{s:sample}.
The data and data analysis methods used in this study are described
in \S \ref{s:data}.
The results of the analysis are presented in \S \ref{s:results}
including analysis of the sample completeness,
a census of ULX candidates, some of their 
properties with respect to their host galaxies, and 
their luminosity distribution function. 
\S \ref{s:discussion} discusses the results.

\section{Sample Definition} \label{s:sample}

We began with a list of all galaxies within 14.5~Mpc 
 contained in the Uppsala Galaxy Catalog (UGC) with photographic magnitude
 within the Catalog's completeness limit of $m_p<14.5$~mag. 
The UGC (Nilson 1973)
 contains all galaxies north of B1950 $\delta=-2^{\circ}30^{\prime}$
 in both of
 two complete samples: galaxies with angular diameters $>$1$^{\prime}$ on
 the first POSS blue prints and galaxies brighter than
 $m_p=14.5$~mag in the Zwicky Catalog of Galaxies and Clusters of Galaxies.
The distance restriction was chosen to insure that any 
 additional \cxo\ X-ray observations needed to complete the survey 
 require integration times of $\le$5~ks in order to detect 20 counts 
($>$3$\sigma$ S/N) in the 0.3$-$6.0~keV \cxo\ band
from a source with a luminosity of $10^{39}$~\ergl\ at the distance 
 of the target galaxy. 
This resulted in a sample of 266 galaxies. 
We then eliminated those galaxies below the completeness limit of the 
{\sl Infrared Astronomical Satellite (IRAS)} 
 survey, $\sim$1.5~Jy at 60~$\mu$m 
(Beichman \etal\ 1988).
This eliminates galaxies with low star-formation rates which are known to
 host few, if any, ULXs (e.g., Grimm \etal\ 2003; Ranalli \etal\ 2003)
The final sample consists of 127 galaxies.

This sample was compared to a compilation extracted from 
 the NASA/IPAC Extragalactic Database (NED)
of all objects designated as a galaxy with 
Virgo-infall velocity $\le 1060$~km~s$^{-1}$
(corresponding to $\sim$14.5~Mpc for $H_o = 73$~km~s$^{-1}$),
and with declination above $\delta=-2.5\arcdeg$ (B1950). This compilation 
contained about 760 objects of which 173 met the UGC and {\sl IRAS} constraints 
imposed on angular size, optical magnitude and far-IR flux. 
The differences between this compilation and our sample
 are ascribed to differences in the distance estimates
(see \S \ref{s:galprops}).
For reference, of the 760 objects,
 579 are below the {\sl IRAS} flux completeness criterion,
 364 fail the UGC photographic completeness criterion,
 and 34 are smaller than 1\arcmin\ diameter.

\section{Properties of ULX Candidates \& Ancillary Data} \label{s:data}

Most of the galaxies in the present sample have been observed with 
 the \cxo\ X-ray Observatory Advanced CCD Imaging Spectrometer (ACIS).
For consistency, \cxo/ACIS data has been used for analysis when available unless a
much deeper \xmm\ observation exists (even then, \cxo\ data were used
 to locate X-ray sources if the observation was judged deep enough to detect 
 ULXs at $>$3$\sigma$ significance). 
For galaxies with multiple exposures,
 the deepest exposure imaging the full optical extent of the target galaxy
 was analyzed unless the target was previously analyzed by 
 Swartz \etal\ (2004, hereafter S04) in which case our original work was used.

About 60\% of the sample galaxies have long-exposure X-ray images adequate to
 detect point-like sources much fainter than ULXs. 
Short observations were made of the remaining galaxies in the 
 survey
 using the \cxo/ACIS-S detector operating in faint, timed-exposure,
 mode. 
These observations were 
 sufficient to {\sl detect} all ULX candidates
 present in the target host galaxies and to accurately 
 determine their positions 
 but not deep enough to constrain their X-ray spectral energy
 distributions in a statistically-meaningful 
 way.\footnote{Superseding deeper observations of a few galaxies have become publicly available since this survey was initiated. They have been used in our analysis instead of our short exposures.}

\subsection{Host Galaxy Properties} \label{s:galprops}

Galaxy morphological type, major isophotal angular diameter
 (measured at surface brightness level
 25.0 mag sec$^{-2}$ in blue light), major-to-minor isophotal diameter ratio,
 and position angle 
 were taken from the Third Reference Catalogue of Bright Galaxies
 (de~Vaucouleurs~\etal\ 1991, hereafter RC3).
These parameters were used to define a source search region, 
 the ``elliptical \DTF-isophote'', for each survey galaxy. 
Source  search regions were overlaid on Digitized Sky Survey blue 
 images of each galaxy to confirm their accuracy. In a few cases,
 the PA or other parameter was clearly in error and adjustments 
 were accordingly made to the region definition.
Galaxy center coordinates were taken from the NED.

\begin{center}
\includegraphics[angle=-90,width=0.47\textwidth]{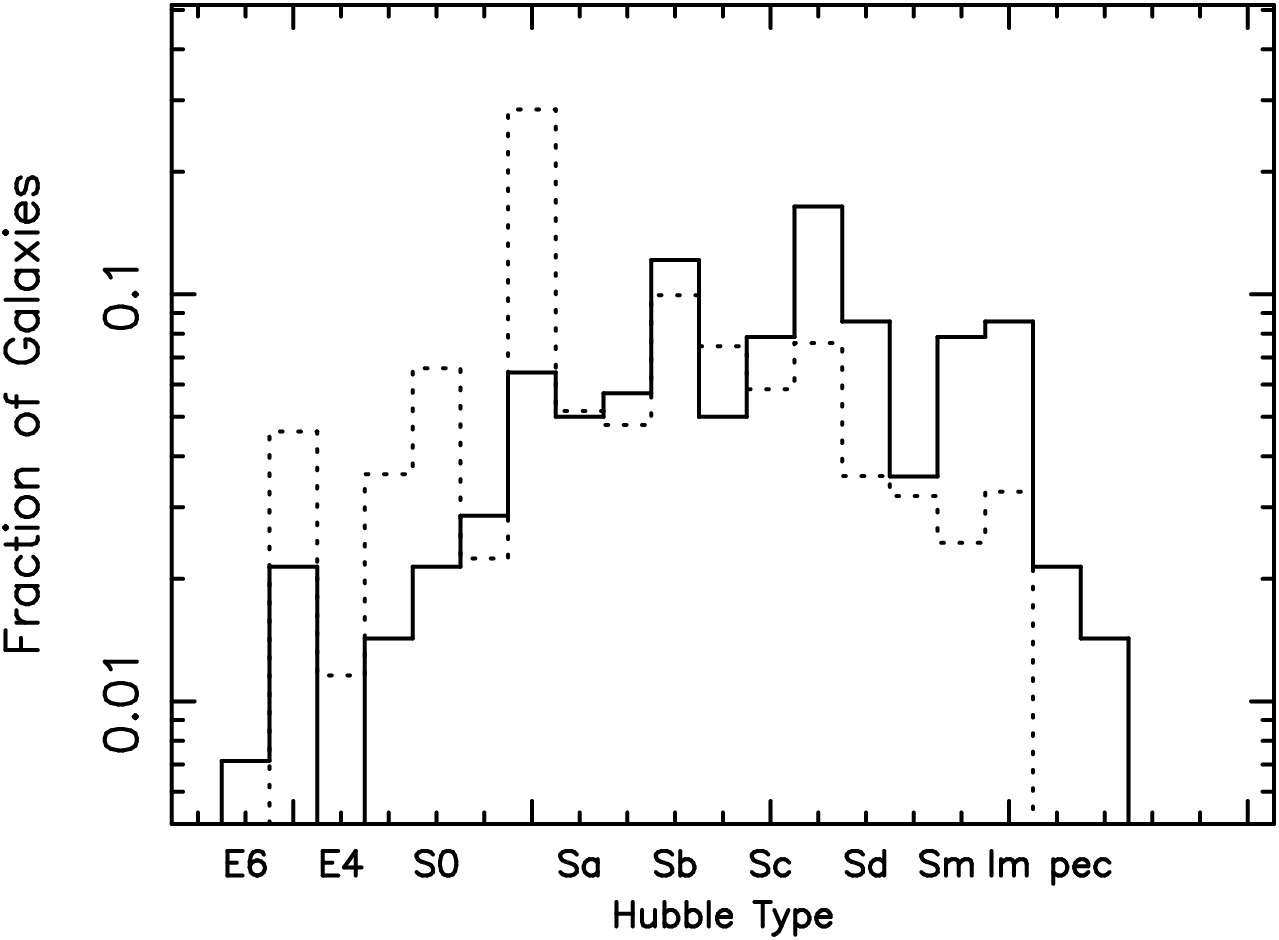} 
\figcaption{Revised Hubble morphological type, $T$, of the present sample of galaxies ({\sl solid histogram})
and of all galaxies in the 3rd Revised Catalog (de~Vauceuleurs \etal\ 1991; {\sl dotted histogram}).
\label{f:ttype}}
\end{center}

Figure~\ref{f:ttype} displays the distribution of the Hubble morphological 
 type of the sample galaxies and of all the galaxies included in the RC3
 with tabulated morphological types (12520 objects). The current sample has
a larger fraction of late-type galaxies and smaller fraction of early-type 
galaxies than the RC3 catalog. This reflects the true scarcity of ellipticals
in the local volume. 
(Note that galaxies of uncertain type are cataloged as $T=0$ galaxies
 which explains the large number of these types in the full RC3.)

Distances to target galaxies were taken from the critically-assessed values 
 used previously in the ULX catalog of S04.
Otherwise, distances were taken from the recent NED compilation
of redshift-independent distances\footnote{http://nedwww.ipac.caltech.edu/Library/Distances/} 
 in the following order of preference:
Cepheid period-luminosity relation from the \hst\ Key Project results
(e.g., Freedman \etal\ 2001), other Cepheid-based
distances, surface brightness fluctuations, the tip of the red giant branch, 
 globular cluster luminosity functions, brightest stars,
and the Tully-Fisher relation. 
Targets not included in the redshift-independent 
NED compilation or in S04 were assigned distances as
listed in the Tully Nearby Galaxies Catalog (Tully 1988)
or, finally, from flow-corrected redshift distances listed in NED,
preferring the values corrected for Local Group motions
(e.g., Karachentsev \& Makarov 1996) over those using only a
single-attractor Virgocentric flow model (e.g., Tully \& Shaya 1984, see also
Mould \etal\ 2000). In the following,
updated distances are used to scale previous results~--~particularly
ULX candidate luminosities.

Total gravitational 
 masses of the sample galaxies were computed from total (asymptotic)
 $B$ magnitudes corrected for 
 inclination and for Galactic and internal extinction (B$_T^0$; RC3) 
 using the scaling  
$\log(M/M_{\odot})=-0.4M_B+2.65$
and distance moduli computed using the distances discussed above.
This scaling is the best linear model fit to the 313 galaxies in the 
Catalog of Nearby Galaxies 
 (CNG; Karachentsev \etal\ 2004, hereafter K04) 
 with tabulated masses based on H{\sc i} rotational velocity curves. 
This scaling gives a $M/L$ ratio of 2.9 \msun/$L_{\odot}$. 
Fifty-three of our sample galaxies are included in the subset of CNG 
galaxies with tabulated masses. A comparison of that tabulation with the 
above expression shows variations not exceeding 0.5 dex in the logarithm
 for sample galaxies within this subset.

Galaxy-wide star-formation rates (SFRs) were estimated from the {\sl IRAS}
60 and 100~$\mu$m flux densities using the conversion given in 
Kennicutt (1998), 
${\rm SFR} ({\rm M}_{\odot}/{\rm yr}) = 0.045(L_{{\rm FIR}}/10^{42})$,
using $L_{\rm{FIR}} = 4 \pi D^2 (1.26\times 10^{-11} (2.58 S_{60} + S_{100}))$
where $S_{60}$ and $S_{100}$ are the 60 and 100~$\mu$m flux densities,
respectively (Helou \etal\ 1985) and $D$ is the distance to the galaxy. 
The $L_{\rm {FIR}} - \rm{SFR}$ relation used here strictly applies to $L_{\rm FIR}$
computed over the broader range 8$\mu$m to 1000$\mu$m which is about $\epsilon \sim 1.7$ times 
higher (e.g., Rowan-Robinson \etal\ 1997)
than the value estimated here using the standard {\sl IRAS} relation from Helou \etal\ (1985).

\subsection{X-Ray Data Reduction}

For \cxo\ observations,
  ACIS Level 2 event lists were generated
  by applying a charge transfer inefficiency (CTI) correction to
 and selecting only standard grades and {\tt status=0} events
   from Level~1 data using the 
 Chandra Interactive Analysis of Observations (CIAO, version 4.2) 
 software tool {\tt acis\_process\_events}.
For \xmm\ observations, 
 event lists were first calibrated from raw EPIC~pn Observation Data Files
  using Science Analysis System v8.0.0 tool {\tt epchain}
 to apply gain and CTI corrections.
Only events with FLAG=0 and PATTERN$\le$4 
 were used in the analysis.
For \ros\ observations, HRI or PSPC 
  basic science data event lists were used.
In all cases, only
 events located within the \elip\ isophote were examined in the analysis.
An observation identifier is listed in column~6 of Table~1.

\subsection{Source Detections}

X-ray sources were located and
 their celestial coordinates, detected counts and estimated background
 within an $\sim$97\% encircled energy radius were tabulated.
Details of the source finding algorithm are given in Tennant (2006).
Source-finding was applied to events in the 
 0.3$-$6.0 keV range for \cxo\ observations, in the
 0.5$-$4.5 keV range for \xmm\ observations, and in the full
 0.1$-$2.4 keV range for \ros\ observations. The different ranges were
 chosen as those that tend to maximize $S/N$.
The duration of the observations were taken from the good-time intervals
 provided as part of the calibrated X-ray event lists.
Binned X-ray light curves spanning the duration of the 
 individual observations and covering most or all of the 
 elliptical \DTF-isophote search region, excluding detected sources,
 were examined for periods of high background. These time periods were 
 excluded, GTIs re-computed, and source-finding repeated.
The aperture-corrected number of source counts 
 (within the detection bandpasses) are listed in column~7 
of Table~1. The $S/N$ ratios are listed in column~8.

\subsection{Source Positions}

The vast majority of the ULX candidates identified in this work
 have been imaged with \cxo's High Resolution Mirror Assembly.
The accuracy of absolute positions in the \cxo\ data
 have a typical rms systematic error of $\sim$0.\arcsec 1 due to
 uncertainties in the plate scale.\footnote{http://asc.harvard.edu/cal/hrma/optaxis/platescale/}
Refined estimates of the centroids of the brighter sources were made by fitting
 an elliptical Gaussian to the spatial distribution of X-ray
 events.
The estimated 
 statistical uncertainty in the source positions due to centroiding errors
 is less than 0.\arcsec 2 for \cxo\ observations for a combined 
 single-axis rms error of $\sim$0.\arcsec 3.
The positional accuracy of sources detected in \xmm\ 
 observations is roughly estimated to be $\sim$1.5\arcsec,
 those in \ros\ observations $\sim$5\arcsec, and 
 the refined estimates are correspondingly larger than those of \cxo-imaged
 sources.
Source celestial positions quoted in this work are these centroid-refined
 positions. They are listed in columns 1 and 2 of Table~1 and the source's
 putative host galaxy and distance are given in columns 3 and 4, respectively.

The radial position of each ULX candidate relative to its host galaxy center,
 expressed as the fraction, $f_{\rm D25}$, of the deprojected galaxy radius in 
 units of $1/2$ the major isophotal diameter, is listed in 
 column 5 of Table~1.

\subsection{Source Luminosities}

ULX candidate luminosities were estimated in two ways.
First, for all sources, a simple conversion from aperture-corrected 
 observed source counts (in the energy ranges defined above)
 to luminosities in the 0.3$-$10.0~keV range was made using the 
 Portable Interactive Multi-Mission Simulator (PIMMS; Mukai 1993).
PIMMS estimated the instrument-specific {\sl unabsorbed} (i.e., intrinsic)
 flux provided the observed source
 count rate and a model for the spectral energy distribution.
An absorbed power law model was assumed in all cases with an absorption 
 column equal to the line-of-sight Galactic column taken from the
H{\sc i} maps of Kalberla \etal\ (2005)
 accessed through the FTOOL utility {\tt nh}
 and a power law index fixed at 1.8 following S04.

Second, for sources with $>$130 counts detected in any observation,
 either spectral fits were made or published results of spectral 
 fits were taken from the literature.
For results taken from the literature, 
 the intrinsic luminosities were converted from the 
 originally-published energy range to the 0.3$-$10.0~keV bandpass
 and, where needed, scaled to the distances adopted here.
PIMMS was used for the bandpass conversion and to scale 
 from observed to intrinsic (unabsorbed) fluxes where needed provided
 spectral model fit parameters were provided.
References to published spectral modeling results 
 are listed in column~11 of Table~1.

For sources with $>$130 detected source counts
 but without previously-published luminosities,
  spectral analysis was performed using the XSPEC v.11.3.2 
 spectral-fitting package with 
 redistribution matrices and ancillary response files appropriate
 to the specific observation.
Absorbed (XSPEC's {\tt phabs}) power law ({\tt powerlaw}), 
 blackbody accretion disk ({\tt diskbb}), 
 and optically-thin diffuse gas ({\tt apec}) models
 were applied, sequentially, to the observed spectra 
 until an acceptable fit was obtained
 according to the $\chi^2$ statistic
(events were grouped to ensure a minimum of 10 counts per spectral energy bin).
The effects of pile-up of sources with detected count
 rates in excess of 0.1 counts~frame$^{-1}$ in \cxo/ACIS observations 
 were treated following the procedures
  described by Davis (2001) using the multiplicative XSPEC model {\tt pileup}.
All ULX candidates in this category, with one exception,
 were fit satisfactorily with an
 absorbed power law model.
The exception is the ULX candidate in NGC~4395 for which no 
single-component model
 provided an acceptable fit to the data. The results of the 
 best-fitting ($\chi^2_{\nu}=1.16$ for 8 dof) {\tt diskbb} model
 are quoted in Table~1 for this object.
Luminosities were then estimated from the model-predicted unabsorbed 
 flux in the 0.3$-$10.0~keV energy band and the adopted distance 
 to the ULX candidate host galaxy.

Luminosities estimated from the observed count rates are listed
 in column~9 of Table~1 and those estimated from spectral fits or taken from
 the literature are listed in column~10.

\section{Results of the ULX Population Census} \label{s:results}

\subsection{Sample Completeness}

ULXs are known to correlate with star formation rate in late-type galaxies
 and with stellar mass in early-type galaxies (e.g., S04).
How well the current sample of galaxies, and the ULX population it
 provides, represents a statistically complete inventory 
can be estimated by measuring the mass and star formation rate of the sample.

\begin{center}
\includegraphics[angle=-90,width=0.47\textwidth]{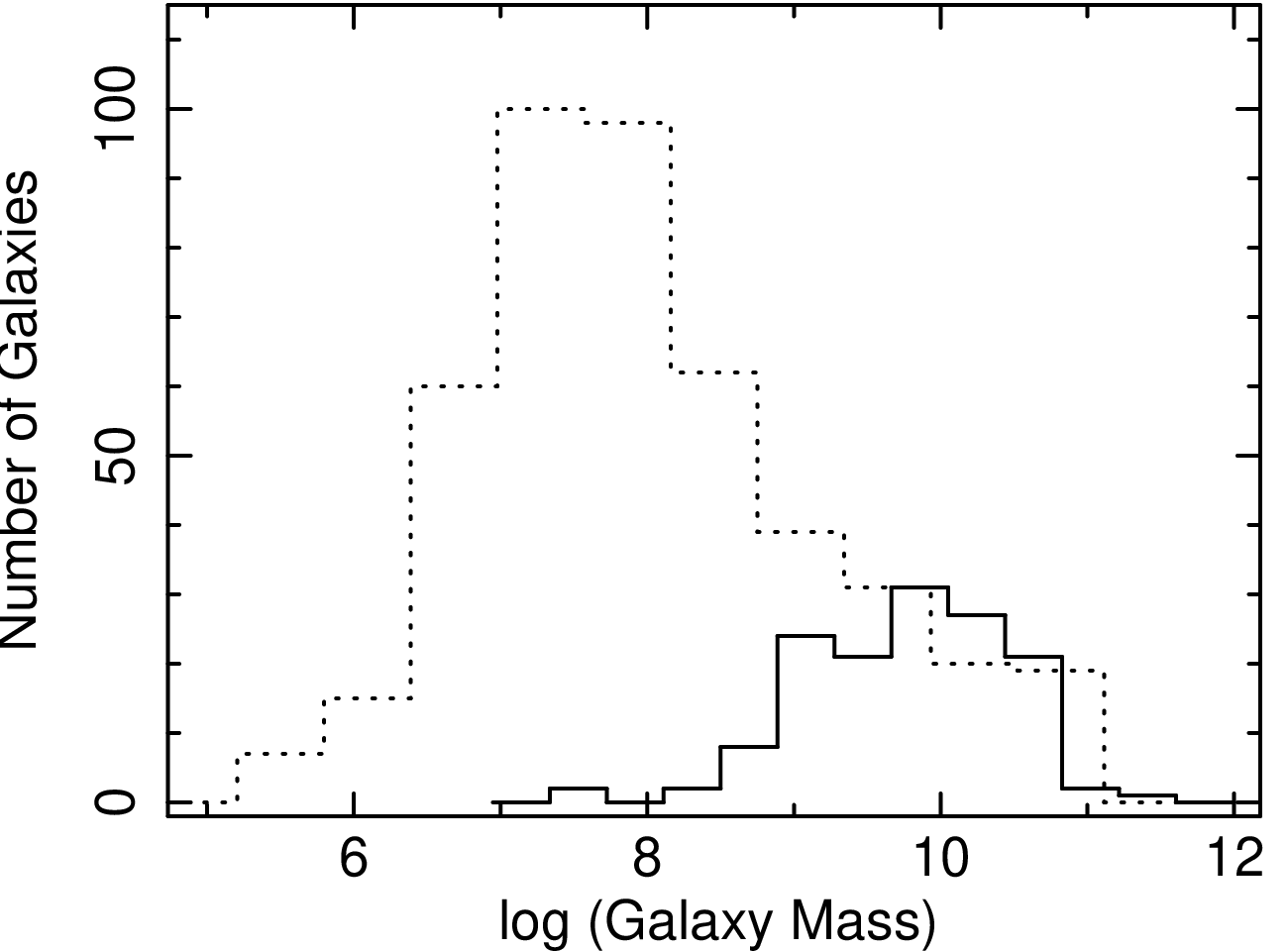} 
\figcaption{Distribution of estimated masses of galaxies in the present sample
({\sl solid histogram}) and of all galaxies in the Catalog of
Neighboring Galaxies (Karachentsev \etal\ 2004; {\sl dotted histogram}).
Masses were estimated from absolute $B$ magnitudes as
$\log(M/M_{\odot})=-0.4M_B+2.65$ (see text).
\label{f:galmass}}
\end{center}

The total dynamical mass in the sample galaxies is
 estimated, by the methods described in \S\ref{s:galprops}, to be
 3.5$\times$10$^{12}$~\msun\ for a mass density of
 5.7$\times$10$^{8}$~\msun~Mpc$^{-3}$ within the sample volume of 6100~Mpc$^{-3}$.
This is slightly less than the density in the Local Volume,  
 7.3$\times$10$^{8}$~\msun~Mpc$^{-3}$ within 8~Mpc (K04),
and slightly more than the mean density measured out to redshifts $z\sim0.2$ 
 (e.g., Blanton \etal\ 2003; Liske \etal\ 2003),
 (3.9-4.2)$\times$10$^{8}$~\msun~Mpc$^{-3}$, assuming $H_o = 73$~km~s$^{-1}$~Mpc$^{-1}$
  and $M/L=2.9$ (see \S\ref{s:galprops}).
We attribute most of the differences to local variations within our small survey volume
 and conclude the current survey accounts for most of the mass within the sample volume.

There is, however, a sizable population of nearby dwarf galaxies 
 that are excluded by our selection criteria.
 Figure~\ref{f:galmass} compares the masses of galaxies in the current sample
to the masses of all galaxies in the CNG (\S \ref{s:galprops}).
The CNG is purported to be complete to roughly $B\sim17$~mag within a 
 distance of $\sim8$~Mpc.
The CNG contains a large number of galaxies with masses below about 
10$^9$~\msun\ compared to the sample of galaxies used here. 
For reference,
 K04 define dwarfs as galaxies with $M_B>-17$ corresponding to masses $M$\LA10$^{9.5}$~\msun.
However, only about 7$\times$10$^{10}$~\msun, or about 4\%, of the total mass in the CNG compilation is 
 in dwarf galaxies. A similar mass percentage is likely omitted in the current survey.

The total SFR in the sample galaxies is 30.3$\epsilon$~\msfr\ based on their FIR 
luminosities\footnote{Here and elsewhere we exclude the SFR contributions from the Seyfert galaxies NGC~1068,
 NGC~660, and NGC~4826 because their high FIR luminosities are likely dominated by their 
 nuclear activity rather than by star formation.} 
 corresponding to a SFR density of 0.005$\epsilon$~\msfr~Mpc$^{-3}$.
As noted in \S \ref{s:data}, these values should be scaled upwards by a factor of $\epsilon \sim1.7$ to 
 account for the full FIR bandpass. 
The resulting density is then within the  range estimated 
from \ha\  measurements (0.013$\pm$0.006~\msfr~Mpc$^{-3}$) by Gallego \etal\ (1995) 
 and from the  [O{\sc ii}] star-formation rate 
 indicator (0.010$\pm$0.005~\msfr~Mpc$^{-3}$, Gallego \etal\ 2002)
 and just below the \ha\ estimates by Hanish \etal\ (2006) of
 $0.011^{+0.003}_{-0.002}$~\msfr~Mpc$^{-3}$ for nearby galaxies.
However, this rate is much less than recent estimates at higher redshifts
 (e.g., 0.025 \msfr~Mpc$^{-3}$, Bothwell \etal\ 2011; 
 0.024~\msfr~Mpc$^{-3}$, Robotham \& Driver 2011 and references therein). 

Our selection criteria were designed to sample most major star-forming galaxies but 
excludes most dwarf galaxies. 
K04 note that the total H{\sc i} mass in
dwarf galaxies is about 15\% of the total amount in their compilation.
If H{\sc i} mass is taken as roughly proportional to SFR (e.g., Kennicutt 1998), 
 then a portion of the total SFR may have been missed in our sample.
Swartz \etal\ (2008) noted that, although ULXs are rare in dwarf galaxies, they do occur at 
a rate proportional to the SFR and that the SFR per unit mass in dwarf galaxies is higher 
than in giant spirals. 

In summary, the current sample of nearby galaxies is reasonably complete
 in both mass and SFR. 
The population undersampled in the survey are the low-mass dwarf galaxies which
 dominate by number but contribute only a small fraction to the total SFR and an even
 smaller fraction of the total mass within the sample volume.
The potential number of ULXs that may be present in this undersampled population
 is estimated in the following section.

\subsection{The Local ULX Census}

There are 107 ULX candidates identified within the current sample of 127 galaxies.
Basic properties of these ULX candidates are listed in Table~1.
There are thus 0.0175 ULX/Mpc$^3$ in the local universe 
 before correcting for background interlopers and sample incompleteness.
From the total estimated mass and SFR for the sample we derive a ULX rate of 
 1 ULX per 3.2$\times$10$^{10}$~\msun\ and 1 ULX per 0.28$\epsilon$ \msfr\ 
 SFR (where $\epsilon \sim 1.7$ is the scaling to the full FIR bandpass).
Fifty-five of the 127 sample galaxies host ULXs.
 Galaxies hosting ULXs account for  71\%
of the total mass in the sample and 
83\% of the total SFR.
If we assume 85\% (following K04) of the actual number of galaxies in the sample volume are 
 dwarf galaxies excluded in the current sample, then there are 850 galaxies in the sample volume
 and the number of galaxies  hosting ULXs is 6.5\% of the total by number.

The number of potential background sources included in the 
 search fields can be estimated from the published fits to the 2.0$-$10.0~keV
 cosmic X-ray background $\log N - \log S$ distribution (Moretti \etal\ 2003)
 as follows. 
If the 2.0$-$10.0~keV flux from a ULX of luminosity $10^{39}$~\ergl\ in a
 galaxy is $S$ (a function of the adopted distance)
 and the total search area of the galaxy is $A$,
 then the expected number of background sources above this flux value is
 $N(>S)=A/(3.26S^{1.57}+0.0073S^{0.44})$.
Applying this to all galaxies in the present sample results in an
estimated 14 background sources. This is a conservative estimate in
that the actual cumulative luminosity function for ULXs (\S \ref{s:xlf}) is 
much flatter than the $\sim S^{-1.57}$ that dominates the 
 background $\log N - \log S$ distribution at fluxes typical of the present
study. In fact, by this same analysis, only 0.4, or about 4\%, of the ULX candidates
 more luminous than 10$^{40}$~\ergl\ are expected to be background interlopers.

\begin{center}
\includegraphics[angle=-90,width=0.47\textwidth]{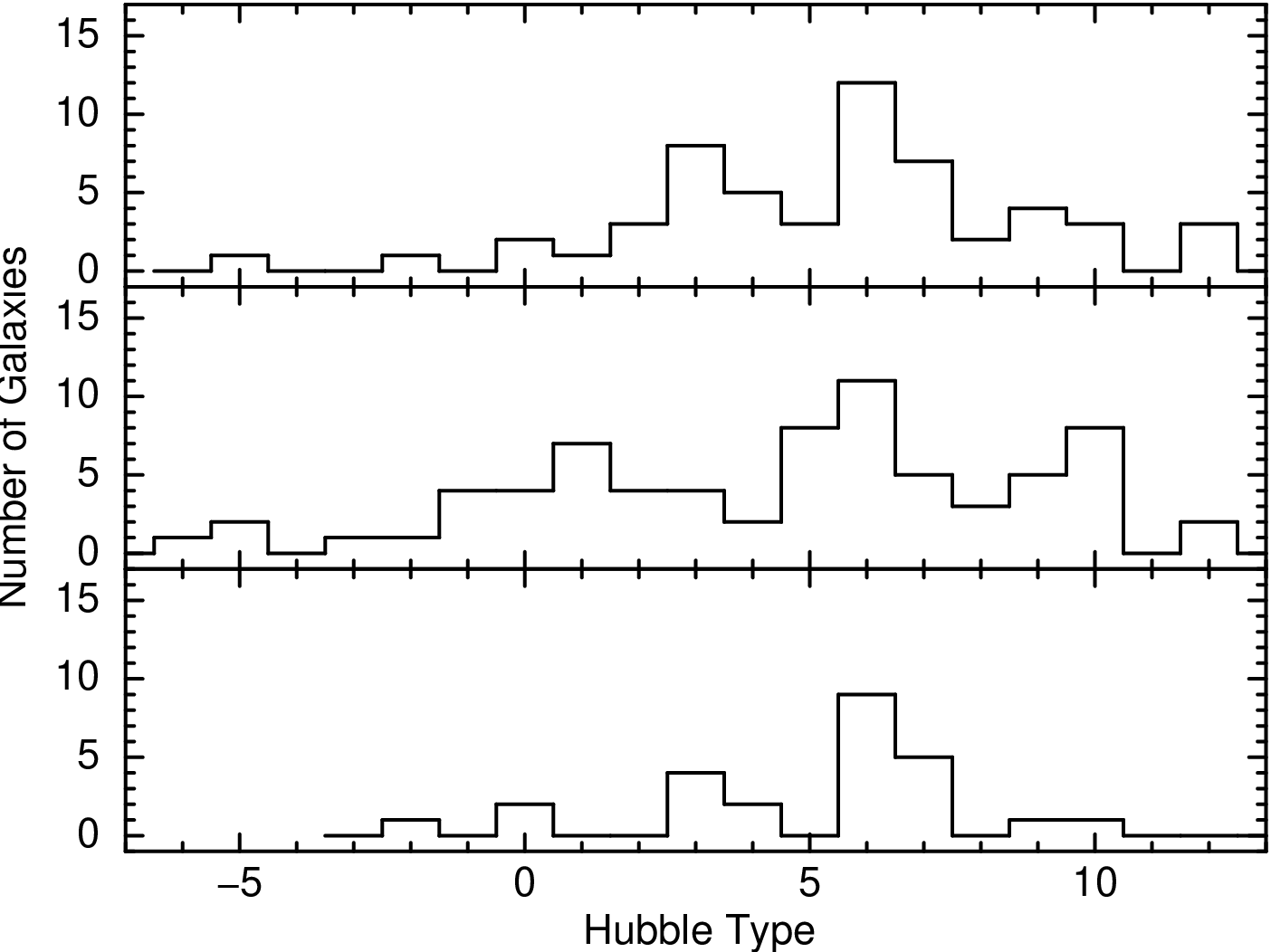} 
\figcaption{Distribution of revised Hubble types for ({\sl top panel}) galaxies hosting ULX candidates,
({\sl middle panel}) galaxies without ULX candidates and ({\sl bottom panel}) galaxies hosting ULXs
with estimated luminosities in excess of 3$\times$10$^{39}$~\ergl.
\label{f:ttype2}}
\end{center}

The number of potential ULX candidates within the sample volume hosted by dwarf galaxies,
 and therefore excluded by our search criteria, can be estimated from the detailed work of 
 Karachentsev \etal\ (K04) on the Local Volume galaxy population.
Assuming the number of ULXs scales with SFR (e.g., Grimm \etal\ 2003; S04) and that the SFR 
scales with the \hi\ mass 
 (Kennicutt 1998; but see also, e.g., Kennicutt \etal\ 2007; Bigiel \etal\ 2008), 
 then we expect an additional 19 ULXs within the sample volume hosted 
by dwarf galaxies because the fraction of \hi\ mass contained in dwarf galaxies is 15\% according to K04.
Assuming the additional ULXs are distributed among 19 individual dwarf
 galaxies, then the number of galaxies 
 hosting ULXs (including dwarfs) rises to 
 8.7\% of the total number of galaxies ($\sim$850) in the sample volume.
The number of additional ULXs estimated by scaling by mass is less than the uncertainties
 that can be ascribed to random fluctuations since only 4\% of the mass is contained in dwarf galaxies (K04).

As with previous surveys, ULX candidates are found in all galaxy morphological types.
Figure~\ref{f:ttype2} displays the distribution of revised Hubble types of galaxies in the sample with no ULXs,
 those hosting ULXs, and those hosting ULXs more luminous than 3$\times$10$^{39}$~\ergl.
The latter catagory typically excludes ULXs in elliptical galaxies (Irwin \etal\ 2004; S04).
There are no statistically-significant differences among the Hubble type distributions of 
galaxies with and without ULXs. There are few early-type galaxies hosting ULXs but there 
are also few early-type galaxies in the sample (cf. also Figure~\ref{f:ttype}). 
There are more ULXs per late-type ($T>0$) sample galaxy, 0.48,  compared to
 the rate in ellipticals, 0.23, but this is only a 2$\sigma$ decrease in the ellipticals 
 compared to the expected value based on the spirals (3 observed ULXs and 6.2$\pm$2.5
 expected in 13 $T\le0$ galaxies).
There is no difference between the number of ULXs per galaxy in early-type spirals
 ($1<T<4$) and in late-type spirals in our sample.
Recently, Walton \etal\ (2011b) found a slight preference for ULXs in early-type spirals
 but only at the 1$\sigma$ level. Walton \etal\ (2011b) did find a higher incidence of ULXs
 in spirals, 0.59, compared to our sample. 
This is likely due to the usual bias towards more massive and luminous galaxies
 inherent in X-ray-selected galaxy samples.

\begin{center}
\includegraphics[angle=-90,width=0.47\textwidth]{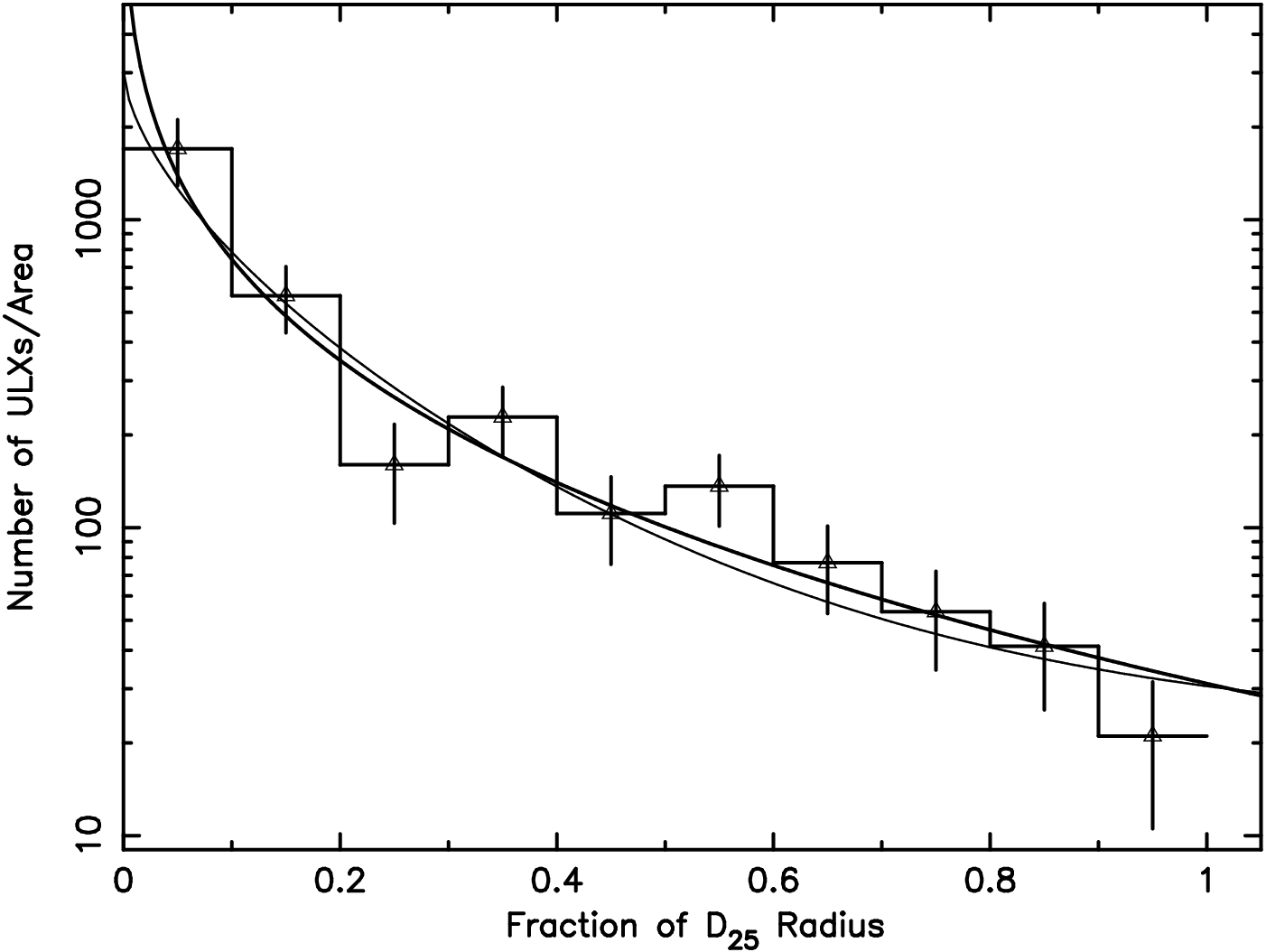} 
\figcaption{Surface distribution of ULX candidates. The abscissa is the deprojected radial position, $f_{\rm D25}$ (Table~1), expressed as a fraction of the host galaxy's angular radius ($\equiv 0.5D_{25}$) and the ordinate is the number of ULX candidates per unit area $fdf$, on the range $f$ to $f+df$.
The thick curve is the best-fitting generalized exponential, $A\exp(-f/h)^{1/n}$.
The best-fit is centrally-peaked with a value $h\sim0.001$ and $n\sim3.6$. The thin curve is the best-fitting function with fixed parameters $h\equiv 0.06$ and $n\equiv 1.6$ corresponding to the best-fit values determined by S04 for a larger sample of 3413 discrete X-ray sources; only the model norm was allowed to vary in determining the thin curve.
\label{f:frad}}
\end{center}

Figure~\ref{f:frad} displays the radial distribution of all ULX candidates normalized to their host galaxy
deprojected $D_{25}$ isophote radius.
The distribution was fitted to a generalized exponential function (Sersic profile) of the form
$A\exp^{-(f_{D25}/h)^{(1/n)}}$. 
The best fit parameter values (resulting in the thick curve of Figure~\ref{f:frad}) 
 approximate a De~Vaucouleurs profile, $n=4$.
Fixing the index to $n=4$ results in a best-fitting scale height $h=3.0_{-1.2}^{+2.4}\times 10^{-4}$ 
(90\% confidence for 1 interesting parameter; $\chi^2=6.9$ for 8 dof).
This indicates a much steeper slope at small values of the abscissa than found previously by S04
 for ULXs and for $\sim$3400 fainter sources, $h=0.06\pm0.03$.
Fixing the index to $n=1.6$ and the scale height to $h=0.06$, the best-fitting values found by S04,
 results in the thin curve shown in Figure~\ref{f:frad}.
Five of the twelve ULX candidates in the range $0\le f < 0.1$ are within 5\arcsec\ of their host
 galaxy nuclei and would have been excluded in our previous survey (S04).
No nuclear exclusion region was imposed in the present work.
This accounts for some of the steepening of the curve 
 at small values of the abscissa but overall the two curves are similar.

\begin{center}
\includegraphics[angle=-90,width=0.47\textwidth]{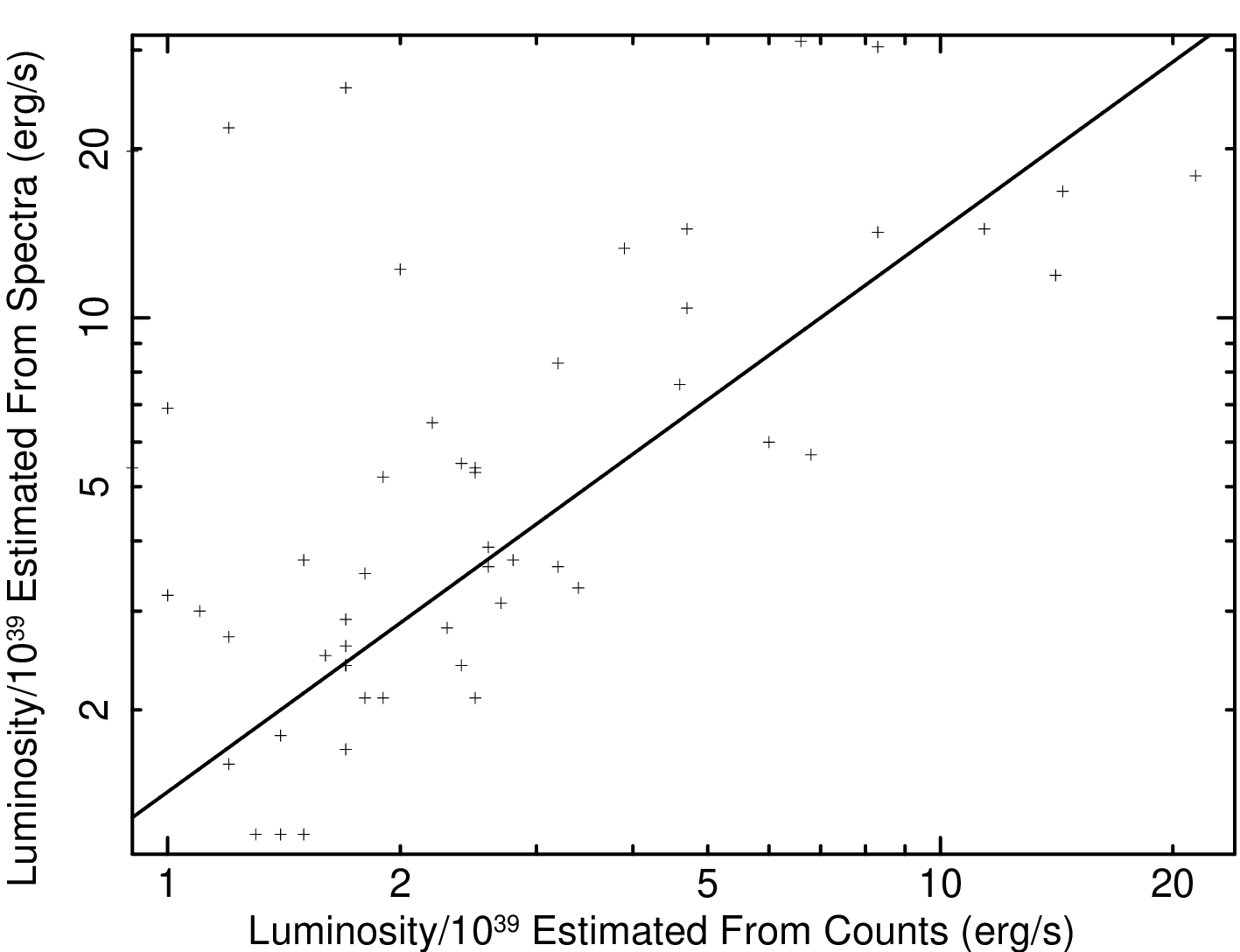} 
\figcaption{Comparison of intrinsic ULX luminosities estimated from spectral fits and from detected counts.
Best-fitting linear function slope is 1.43 showing that luminosities obtained from spectral fits generally
exceeds luminosities estimated by numbers of detected counts. This is primarily attributable to absorbing
column densities estimated by spectral fitting exceeding the line of sight Galactic column densities assumed
in the calculation of luminosities from detected counts. The scatter of the data is due to differences in
fitted spectral shapes (mostly power law indices but also thermal or other components in some cases)
compared to the $\Gamma=1.8$ power law assumed otherwise.
\label{f:LctsvsLsp}}
\end{center}

\begin{center}
\begin{figure*}[t]
\includegraphics[angle=-90,width=0.47\textwidth]{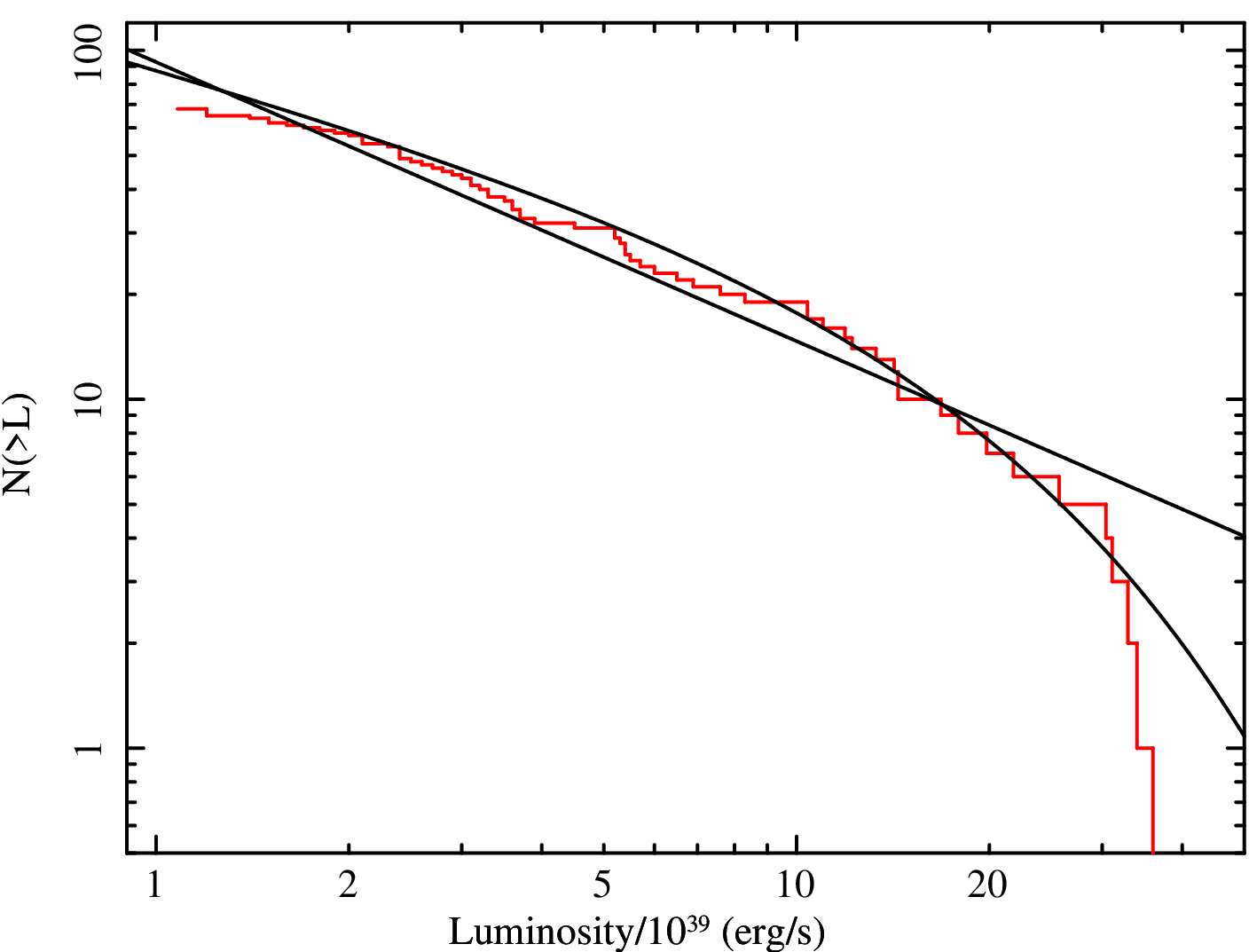} 
\includegraphics[angle=-90,width=0.47\textwidth]{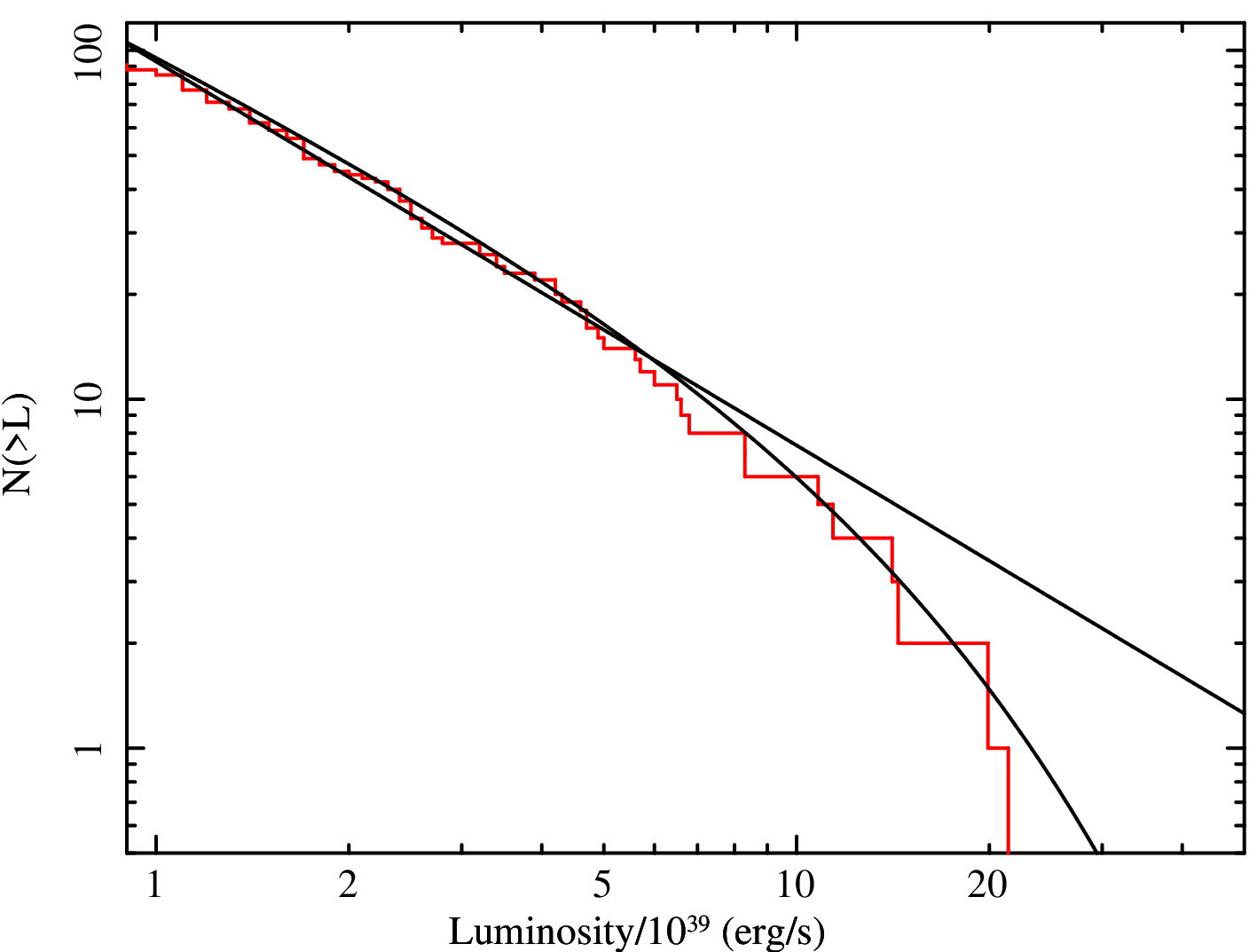} 
\vskip 10pt
\includegraphics[angle=-90,width=0.47\textwidth]{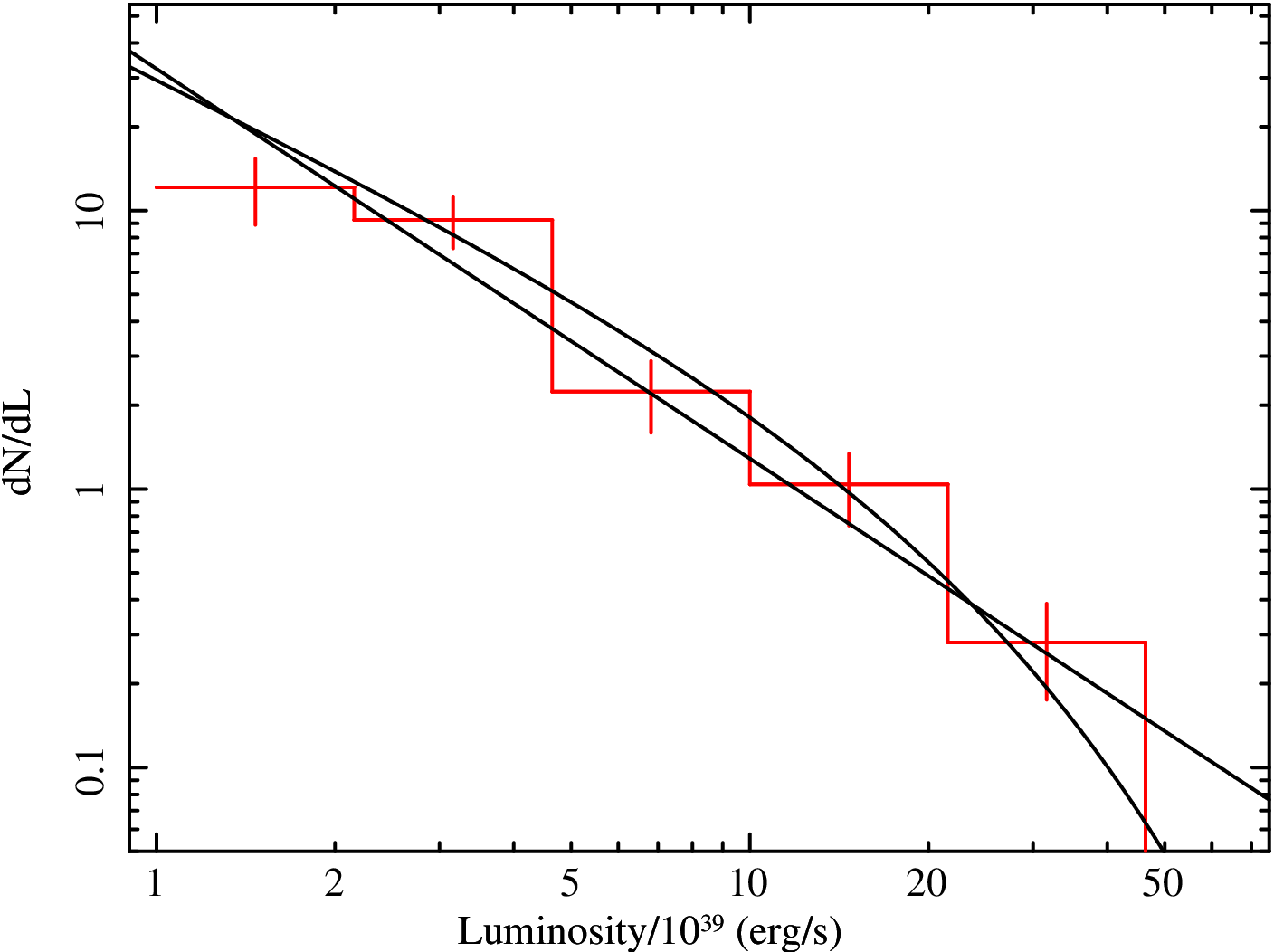} 
\includegraphics[angle=-90,width=0.47\textwidth]{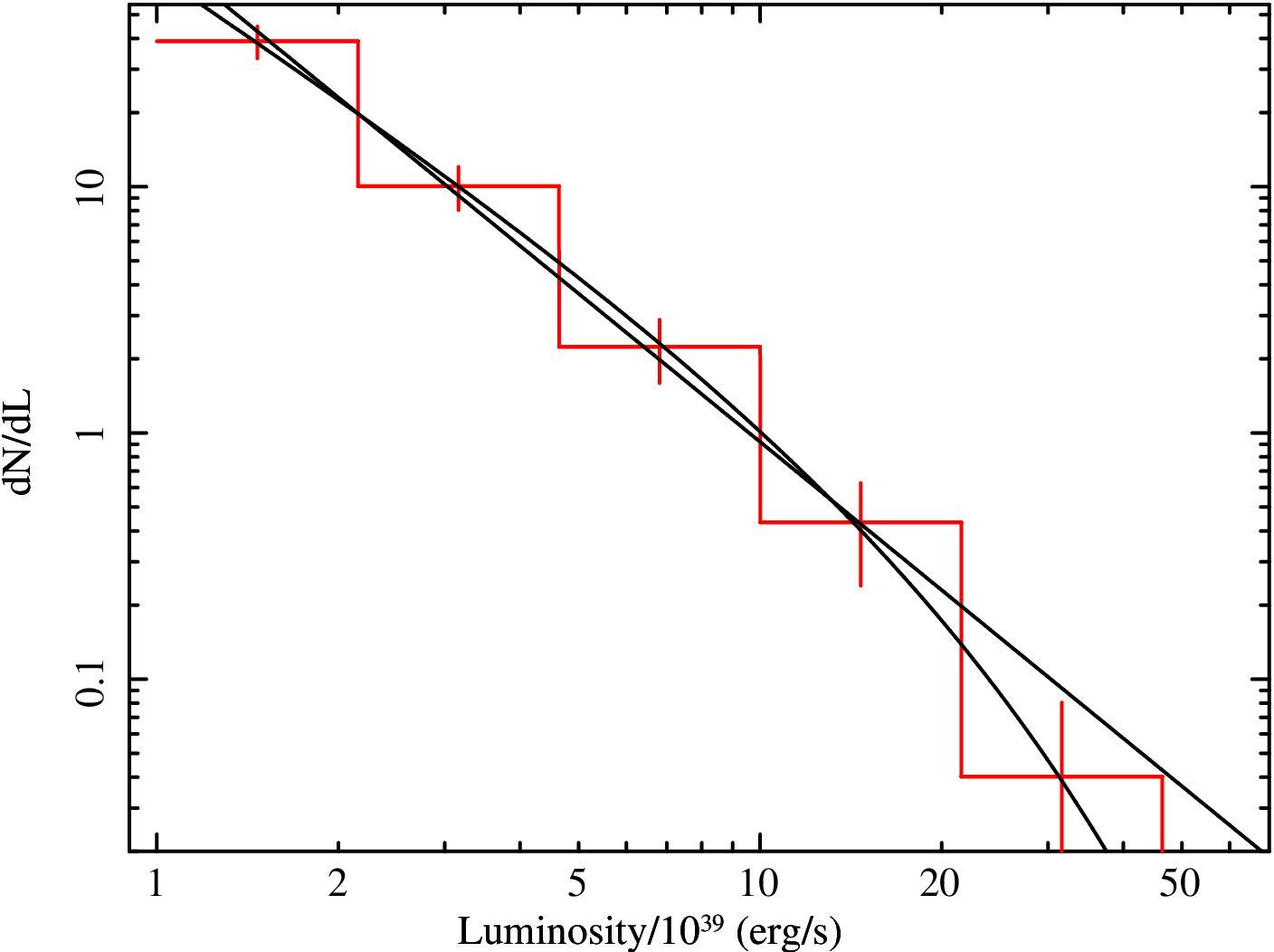} 
\figcaption{Cumulative ({\sl upper panels}) and differential ({\sl lower panels}) ULX luminosity functions
using the luminosities estimated from spectral fitting ({\sl left panels}) and luminosities estimated
from the detected number of counts ({\sl right panels}). Two models were applied in all cases,
a power law with an exponential cutoff, $C L_{\rm X}^{-\alpha_1} \exp(-L_{\rm X}/L_{c})$,
and a 
a pure power law model, $C  L_{\rm X}^{-\alpha_1}$.
See Table~2 for model fit parameters.
\label{f:xlf}}
\end{figure*}
\end{center}

\subsection{The ULX Luminosity Function} \label{s:xlf}

Table~1 lists estimated intrinsic (absorption-corrected)
 ULX luminosities based on the detected number of counts and based on 
 fitting models to the observed spectra.
Figure~\ref{f:LctsvsLsp} shows there are differences between these two 
 estimates. The best linear model fit has a slope of 1.43 meaning that 
 spectroscopically-estimated intrinsic luminosities tend to exceed 
 the values estimated from the number of detected counts.
This is due primarily to the large column densities 
 derived in the spectral fitting which correlates with a higher estimated intrinsic luminosity
 compared to the luminosity based on counts which assumes a total column density 
 equal to the Galactic \hi\ column along the line of sight.
Column densities of ULX candidates were determined to be, on average, about 4 times the 
 line-of-sight Galactic column densities by S04.
There is also an overall scatter in the luminosity comparison shown in Figure~\ref{f:LctsvsLsp}.
This is due to differences between the modeled spectral shape and the simplistic assumption 
 of a $\Gamma=1.8$ power law used to estimate luminosities from the detected numbers of counts.
While spectroscopic luminosities are preferred as the most accurate estimates,
 only a subset of the current sample of ULXs have spectroscopic luminosity estimates
 and the subset is an X-ray-selected subset and hence biased.

With these differences in mind, Figure~\ref{f:xlf} displays the ULX luminosity function. 
The upper panels show the integral or cumulative form, $N(>L)$ against $L$, 
 and the lower panels the differential, $dN/dL$, luminosity functions.
The leftmost panels use spectroscopic luminosities and the rightmost panels
 the counts-based luminosities.
In all cases, two models were fit to the data:
a power law with an exponential cutoff, $C L_{\rm X}^{-\alpha} \exp(-L_{\rm X}/L_{c})$,
and a 
pure power law, $C L_{\rm X}^{-\alpha}$,
 where $\alpha$ is the power law index, $L_{c}$ is the cutoff luminosity, and $C$ is the normalization constant.
Table~2 records the values of the best-fit parameters for the models.
In all cases, a minimum-likelihood fit statistic was used.
The formal uncertainties on the model parameters 
 are not given for the cumulative luminosity functions because the data bins in that case
 are not independent.
However, it is clear from the figure that the pure power law is a very poor representation of the data.
Note the large formal uncertainties in the differential luminosity function fits
 due to the strong correlations among the fit parameters even when the fit statistic is near unity
 indicating an acceptable fit to the data.

The cutoff power law model provides an improvement over the pure power law model in all cases.
The change in the fit statistic, $\ell$, by adding one parameter is 3.46 and 11.19 for the differential 
 luminosity functions based on counts and on spectral fits, respectively, corresponding to 
 4 and 8\% probabilities due to chance alone.

In all cases, the cutoff luminosity is of order $2\times 10^{40}$~\ergl\ and the 
 ($dN/dL$) power law slope is  $\sim$1.6 
 consistent with most values reported previously (e.g., S04; Grimm \etal\ 2003;
 Mineo \etal\ 2011).
Walton \etal\ (2011b) report an acceptable fit (normalized statistic 0.89) from a pure power 
law model applied to a sample of 121 ULX candidates. 
Inspection of their integrated luminosity function, $N(>L)$, displayed in their Figure~7,
 reveals a cutoff power law with a value for the cutoff luminosity similar to the 
 value determined here should also provide an acceptable fit to their data.
A pure power law clearly does not fit our integrated $N(>L)$ luminosity functions 
 (Figure~\ref{f:xlf} upper panels) but is an 
 acceptable fit to our differential $dN/dL$ constructed from the 
 subset of spectrally-determined luminosities
 (Figure~\ref{f:xlf} lower left panel).
 
The differential luminosity function can also be expressed in terms of the 
 SFR. Following Grimm \etal\ (2003), we have, for a combined SFR of 51.4 \msfr\ in our sample
 and a pure power law representation of the differential luminosity function (Table~2),
 $dN/dL_{cts} = 1.79\pm0.48 {\rm SFR} L_{cts}^{-2.0}$ where $L_{cts}$ is in units of 
 $10^{39}$~\ergl. 
Scaling Equation~6 of Grimm \etal\ (2003) to the same luminosity normalization and 
 accounting for the narrower bandpass, 2$-$10~keV, over which they estimated luminosities
 results in an equivalent $dN/dL = 0.6 {\rm SFR} L_{cts}^{-1.6}$ for $L<L_c=21.0$
 which indicates our present census contains roughly twice the number of ULX candidates
 per unit SFR compared to Grimm \etal\ (2003).
However, we note that applying our FIR-based estimate of the SFR (\S \ref{s:galprops})
 to the galaxy sample used to derive Equation~6 of Grimm \etal\ (2003)
 results in a 30\% reduction in the total SFR compared to their adopted value.
This accounts for much of the discrepancy between these two results
 and suggests that they are roughly consistent. 
 
\section{Discussion} \label{s:discussion}

A sample of galaxies 
  with known selection criteria has been constructed in order
   to better quantify the local population of ULX sources and to 
   refine our understanding of the relationships between ULXs,
 their environments, and other X-ray sources. 
We have shown that the sample of galaxies is essentially complete in mass and SFR within 
  a volume of some 6100~Mpc$^3$ and contains 107 ULX candidates.
Within this volume, there is  1 ULX per 3.2$\times$10$^{10}$~\msun\ and 1 ULX per 0.28$\epsilon$ \msfr\ 
 SFR where $\epsilon \sim 1.7$.
 
The ULX number density per unit mass and star formation rate of 
 their host galaxies is consistent with the extrapolation of the 
 luminosity function of ordinary X-ray binaries (Grimm \etal\ 2003; Gilfanov 2004).
This suggests that most black hole X-ray binaries with luminosities above and below
 $10^{39}$~\ergl\ originate through similar stellar and binary evolution processes.
The ULX subclass is consistent with being the extreme end of this distribution,
 perhaps in terms of accretion rate and the black hole mass (which can be as high as 
 80~\msun\ in metal-poor environments: Belczynski \etal\ 2010).
If this is the case, then the onset and duration of the bright X-ray phase in
 ULXs would also be subject to the same constraints as in ordinary X-ray binaries
 with the donor star expanding to fill its Roche lobe.
 
Because of the completeness of the current sample, 
 we can extrapolate the results to somewhat larger volumes
 to predict properties of the expected population of ULXs more generally.
Of particular interest are those objects that are missed in the current
 sample because they are too rare to appear within random small volumes, namely the very luminous ULXs
 of which a handful have been recently reported (e.g., Sutton \etal\ 2011; Farrell \etal\ 2009).

As luminosity is the defining characteristic of ULXs, 
 the ULX luminosity function can be used to quantify if, indeed, there exists another class of rare high luminosity 
 objects unrelated to the general ULX population.
We can analytically extrapolate the differential luminosity function from the parameters given in Table~2
 and express the results in terms of the volume (or corresponding radius) within which a ULX of 
 a given luminosity would be expected assuming a uniform spatial distribution of ULXs. 
The cutoff power law model predicts 1 ULX with a luminosity 
 $L_{\rm X}/10^{39} > 50,100, {\rm and \,} 200$ within a radius of
 13$-$29, 42$-$119, and $\gtrsim$300~Mpc where the low  and high end of the quoted
 ranges corresponds to the differential luminosity functions based on spectral fits and on counts, respectively.
This suggests that ULXs with luminosities up to 10$^{41}$~\ergl\ can be expected within a distance of $\sim$100~Mpc
 but that more luminous objects would not be consistent with the population identified here. 
  
Recently, Sutton \etal\ (2011) described 7 high luminosity ULX candidates, $L_{\rm X} > 5\times 10^{40}$~\ergl,
 identified as part of a 
 cross-correlation of the RC3 catalog of galaxies with the 2nd \xmm\ catalog of 
 serendipitously-detected X-ray sources from pointed observations (Watson \etal\ 2009).
Using revised distances according to information available from the NED, three of these objects
have luminosities below 5$\times$10$^{40}$~\ergl, including one included in the present sample.
Sutton \etal\ (2011) estimate the luminosities (and distances) to the remaining ULX candidates as,
 in units of 10$^{39}$~\ergl\ (and Mpc):  70 (96.2), 61 (32.8), 103 (32.7), and 270 (95.1). 
Using the same arguments as above, we expect the numbers of ULXs at (or above) these luminosities
 and at these distances (within a spherical volume)
 to be: 6.8$-$105, 2.4$-$31.8, 0.07$-$01.92, and $\ll$10$^{-4}$, respectively. 
In other words, the existence of the two brightest ULX
 candidates are inconsistent with expectations.
Note that, above our exponential cutoff of a few $10^{40}$~\ergl, background
 sources dominate the candidate source population and this factor is not included in the 
 estimates made in this section.
The actual coverage of the 2XMM catalog is 360~deg$^2$ of the sky (Watson \etal\ 2009). 
If we assume the catalog reaches a uniform depth for the detection of (very luminous) ULXs of 
 $100d_{100}$~Mpc,
 then the effective volume of the 2XMM catalog is $37000 d_{100}^3$~Mpc$^3$.
The corresponding numbers of ULXs expected reduces to 
 0.09$-$1.40, 0.20$-$2.60, 0.006$-$0.16, and $\ll$10$^{-4}$.

Farrell \etal\ (2009) report an object apparently associated with the galaxy ESO~243-49 with a luminosity
 reaching 1.1$\times$10$^{42}$~\ergl. 
Clearly, such a high luminosity is not expected based on the distribution of luminosities reported here.
Instead, we may ask if the host galaxy displays any peculiar properties that would be evident in our sample.
The galaxy is a large type S0/a edge-on lenticular 90~Mpc distant. 
Luminous ULXs are not common in early-type galaxies (Irwin \etal\ 2004; S04; Liu \etal\ 2006).
The galaxy is not detected in {\sl IRAS} images which places a conservative upper limit to its star-formation rate
 of $\sim$3~\msfr\ but,
because it is an early-type galaxy and viewed edge-on (which tends to enhance IR 
 emission related to disk star formation and nuclear activity), its actual star-formation rate is likely
 much lower and therefore unlikely to host luminous ULXs (Grimm \etal\ 2003; S04).
However, it is a large massive galaxy ($M_B = -19.85$; $\log(M/M_{\odot})=10.6$) that 
 would likely be detectable by {\sl IRAS}  were it located within our sample volume and it would
 be well within our other selection criteria for angular size and photographic brightness.
Finally, the ULX candidate is located in the halo of ESO~243-49 which is an unlikely location
 for a ULX (Swartz 2006).
In summary, a new class of object is needed to account for ULXs such as ESO~243-49 HLX-1 and for any
 ULX in excess of $\sim$2$\times$10$^{41}$~\ergl\ in the local universe.


\clearpage
{\tiny
\begin{deluxetable}{cccrrrrrrrc}
\tablecolumns{11}
\tablewidth{0pt}
\tablenum{1} 
\tablecaption{Properties of ULX Candidates} 
\tablehead{
\colhead{R.A.} &  
\colhead{Decl.} & 
\colhead{Galaxy} & 
\colhead{$D$} & 
\colhead{$f_{\rm D25}^a$} & 
\colhead{ObsID} & 
\colhead{cts} & 
\colhead{$S/N$} & 
\colhead{$L_{{\rm cts}}^b/10^{39}$} &
\colhead{$L_{{\rm X}}^c/10^{39}$} &
\colhead{Ref.$^d$} \\
\colhead{(h m s)} & 
\colhead{(d m s)} & 
\colhead{} & 
\colhead{(Mpc)} & 
\colhead{} & 
\colhead{} & 
\colhead{} & \colhead{} 
& \colhead{(erg s$^{-1}$)} &
\colhead{(erg s$^{-1}$)} &
\colhead{}
}
\startdata
 01 36 51.08  &  15 45 46.9 & NGC 628 &  9.7 &  0.54 & 2057 &    354.1 &   16.5 &   0.8\hspace{10pt} & 2.3\hspace{10pt} &  \\ 
 02 14 04.08  &  27 52 39.5 & NGC 855 &  9.7 &  0.11 & 9550 &   1032.6 &   27.4 &   1.9\hspace{10pt} & 2.1\hspace{10pt} &  \\ 
 02 22 31.26  &  42 19 57.8 & NGC 891 & 10.0 &  0.16 & 794 &   1695.7 &   36.4 &   3.9\hspace{10pt} & 6.7\hspace{10pt} &  \\ 
 02 22 31.36  &  42 20 24.4 & NGC 891 & 10.0 &  0.15 & 794 &   2042.7 &   39.0 &   4.7\hspace{10pt} & 10.4\hspace{10pt} &  \\ 
 02 27 21.52  &  33 35 00.7 & NGC 925 &  9.1 &  0.23 & 7104 &     96.2 &    8.6 &   4.2\hspace{10pt} & $-$\hspace{10pt} &  \\ 
 02 27 27.53  &  33 34 43.0 & NGC 925 &  9.1 &  0.44 & 7104 &    458.3 &   18.5 &  19.9\hspace{10pt} & 24.8\hspace{10pt} &  \\ 
 02 36 23.73  &  38 59 08.5 & IC 239 & 14.2 &  0.57 & 7131 &     46.3 &    5.9 &   2.4\hspace{10pt} & $-$\hspace{10pt} &  \\ 
 02 36 25.94  &  38 58 03.6 & IC 239 & 14.2 &  0.20 & 7131 &     31.9 &    5.1 &   1.7\hspace{10pt} & $-$\hspace{10pt} &  \\ 
 02 36 29.75  &  39 00 14.6 & IC 239 & 14.2 &  0.91 & 7131 &     64.9 &    7.2 &   3.4\hspace{10pt} & $-$\hspace{10pt} &  \\ 
 02 39 09.31  &  40 52 33.2 & NGC 1003 & 10.6 &  0.53 & 7116 &     45.1 &    6.0 &   2.3\hspace{10pt} & $-$\hspace{10pt} &  \\ 
 02 39 11.66  &  40 53 13.7 & NGC 1003 & 10.6 &  0.93 & 7116 &     26.5 &    4.7 &   1.3\hspace{10pt} & $-$\hspace{10pt} &  \\ 
 02 39 14.35  &  30 08 54.7 & NGC 1012 & 14.4 &  0.18 & 7133 &    108.8 &    9.2 &   5.7\hspace{10pt} & $-$\hspace{10pt} &  \\ 
 02 39 14.73  &  30 08 35.0 & NGC 1012 & 14.4 &  0.51 & 7133 &     21.6 &    3.8 &   1.1\hspace{10pt} & $-$\hspace{10pt} &  \\ 
 02 41 38.79  &  00 27 36.6 & NGC 1055 & 12.6 &  0.65 & 4011 &     30.5 &    4.9 &   1.1\hspace{10pt} & $-$\hspace{10pt} &  \\ 
 02 42 38.89  & $-$00 00 55.1 & NGC 1068 & 14.4 &  0.56 & 344 &   1407.9 &   32.1 &   6.6\hspace{10pt} & 31.1\hspace{10pt} &  \\ 
 02 42 39.71  & $-$00 01 01.4 & NGC 1068 & 14.4 &  0.59 & 344 &    528.4 &   19.5 &   2.5\hspace{10pt} & 2.1\hspace{10pt} &  \\ 
 02 42 40.43  & $-$00 00 52.6 & NGC 1068 & 14.4 &  0.55 & 344 &    221.4 &    8.6 &   1.0\hspace{10pt} & 6.9\hspace{10pt} &  \\ 
 03 45 55.61  &  68 04 55.3 &  IC 342  &  3.3 &  0.48 & 0206890101 &   $-$ &  $-$ &   $-$\hspace{10pt} & 11.0\hspace{10pt} & 1 \\ 
 03 46 15.64  &  68 11 12.2 &  IC 342  &  3.3 &  0.60 & 0206890101 &   $-$ &  $-$ &   $-$\hspace{10pt} & 36.0\hspace{10pt} & 1 \\ 
 07 36 25.53  &  65 35 40.0 & NGC 2403 &  3.1 &  0.37 & 2014 &   5649.6 &   65.5 &   1.7\hspace{10pt} & 2.4\hspace{10pt} &  \\ 
 08 01 48.10  &  50 43 54.6 & NGC 2500 & 10.1 &  0.66 & 7112 &    110.8 &    9.2 &   5.0\hspace{10pt} & $-$\hspace{10pt} &  \\ 
 08 01 57.85  &  50 43 39.5 & NGC 2500 & 10.1 &  0.69 & 7112 &     23.3 &    4.3 &   1.1\hspace{10pt} & $-$\hspace{10pt} &  \\ 
 08 14 37.02  &  49 03 26.6 & NGC 2541 & 11.2 &  0.35 & 1635 &     54.8 &    6.7 &   4.2\hspace{10pt} & $-$\hspace{10pt} &  \\ 
 08 19 29.00  &  70 42 19.3 & UGC 4305 &  3.4 &  0.69 & 1564 &   2435.7 &   40.9 &   6.0\hspace{10pt} & 6.0\hspace{10pt} &  \\ 
 09 22 02.22  &  50 58 54.2 & NGC 2841 & 13.1 &  0.10 & 6096 &    175.5 &   11.8 &   1.1\hspace{10pt} & 3.0\hspace{10pt} &  \\ 
 09 32 06.16  &  21 30 58.7 & NGC 2903 &  8.9 &  0.37 & 11260 &   3028.8 &   49.0 &   2.8\hspace{10pt} & 3.7\hspace{10pt} &  \\ 
 09 55 32.97  &  69 00 33.4 & NGC 3031 &  3.6 &  0.30 & 735 &   8385.7 &   76.6 &   2.5\hspace{10pt} & 5.4\hspace{10pt} &  \\ 
 09 55 46.45  &  69 40 40.3 & NGC 3034 &  3.9 &  0.10 & 361 &   1165.0 &   29.1 &   0.6\hspace{10pt} & 1.5\hspace{10pt} &  \\ 
 09 55 50.01  &  69 40 46.0 & NGC 3034 &  3.9 &  0.04 & 361 &   3352.3 &   48.2 &   1.7\hspace{10pt} & 25.7\hspace{10pt} &  \\ 
 09 55 50.58  &  69 40 43.2 & NGC 3034 &  3.9 &  0.03 & 361 &    552.0 &   20.5 &   0.3\hspace{10pt} & 1.9\hspace{10pt} &  \\ 
 09 55 51.16  &  69 40 43.4 & NGC 3034 &  3.9 &  0.02 & 361 &   1110.8 &   27.2 &   0.6\hspace{10pt} & 32.9\hspace{10pt} &  \\ 
 09 55 52.13  &  69 40 53.1 & NGC 3034 &  3.9 &  0.04 & 361 &    764.7 &   22.9 &   0.4\hspace{10pt} & 2.4\hspace{10pt} &  \\ 
 09 55 54.56  &  69 41 00.4 & NGC 3034 &  3.9 &  0.07 & 361 &    352.2 &   15.9 &   0.2\hspace{10pt} & 5.2\hspace{10pt} &  \\ 
 10 18 12.05  &  41 24 20.7 & NGC 3184 & 14.4 &  0.42 & 804 &    454.0 &   19.0 &   2.3\hspace{10pt} & 2.8\hspace{10pt} &  \\ 
 10 18 23.00  &  41 27 41.7 & NGC 3184 & 14.4 &  0.72 & 804 &    515.8 &   20.0 &   2.6\hspace{10pt} & 3.6\hspace{10pt} &  \\ 
 10 25 06.98  &  17 09 47.2 & NGC 3239 &  8.1 &  0.30 & 7094 &     31.6 &    4.8 &   1.2\hspace{10pt} & $-$\hspace{10pt} &  \\ 
 10 25 08.20  &  17 09 48.3 & NGC 3239 &  8.1 &  0.48 & 7094 &     43.5 &    5.8 &   1.6\hspace{10pt} & $-$\hspace{10pt} &  \\ 
 11 01 13.67  &  03 36 15.6 & NGC 3495 & 12.8 &  0.66 & 7126 &     29.9 &    4.6 &   1.4\hspace{10pt} & $-$\hspace{10pt} &  \\ 
 11 05 45.62  &  00 00 16.2 & NGC 3521 &  7.2 &  0.57 & 9552 &   2401.9 &   42.4 &   1.9\hspace{10pt} & 5.2\hspace{10pt} &  \\ 
 11 11 26.05  &  55 40 16.8 & NGC 3556 & 14.1 &  0.17 & 2025 &   1346.5 &   32.0 &   4.6\hspace{10pt} & 7.6\hspace{10pt} &  \\ 
 11 11 30.34  &  55 40 31.3 & NGC 3556 & 14.1 &  0.08 & 2025 &    355.7 &   16.7 &   1.2\hspace{10pt} & 2.7\hspace{10pt} &  \\ 
 11 11 41.42  &  55 40 57.8 & NGC 3556 & 14.1 &  0.42 & 2025 &    252.9 &   13.5 &   0.9\hspace{10pt} & 5.4\hspace{10pt} &  \\ 
 11 18 58.55  &  13 05 31.0 & NGC 3623 &  7.3 &  0.45 & 1637 &     30.4 &    4.8 &   1.0\hspace{10pt} & $-$\hspace{10pt} &  \\ 
 11 20 15.76  &  13 35 13.7 & NGC 3628 & 10.0 &  0.87 & 2039 &   2946.7 &   45.6 &  14.1\hspace{10pt} & 11.9\hspace{10pt} &  \\ 
 11 20 18.32  &  12 59 00.8 & NGC 3627 &  9.4 &  0.38 & 9548 &    895.1 &   25.4 &   1.7\hspace{10pt} & 2.6\hspace{10pt} &  \\ 
 11 20 20.90  &  12 58 46.6 & NGC 3627 &  9.4 &  0.68 & 9548 &   6068.5 &   65.7 &  11.4\hspace{10pt} & 14.4\hspace{10pt} &  \\ 
 11 20 37.37  &  13 34 29.2 & NGC 3628 & 10.0 &  0.61 & 2918 &    274.1 &   14.2 &   3.5\hspace{10pt} & 1.7\hspace{10pt} &  2 \\ 
 11 26 07.33  &  43 34 06.3 & NGC 3675 & 12.8 &  0.39 & 396 &     14.0 &    3.2 &   1.4\hspace{10pt} & $-$\hspace{10pt} &  \\ 
 11 52 37.36  & $-$02 28 07.1 & UGC 6850 & 14.5 &  0.16 & 7135 &     34.7 &    5.2 &   1.6\hspace{10pt} & $-$\hspace{10pt} &  \\ 
 12 06 07.98  &  47 30 21.2 & NGC 4096 &  8.8 &  0.84 & 7103 &     22.7 &    4.3 &   1.1\hspace{10pt} & $-$\hspace{10pt} &  \\ 
 12 09 22.18  &  29 55 59.7 & NGC 4136 &  9.7 &  0.55 & 2921 &    352.3 &   16.0 &   1.8\hspace{10pt} & 2.1\hspace{10pt} &  \\ 
 12 10 33.76  &  30 23 58.0 & NGC 4150 & 12.8 &  0.12 & 1638 &     55.7 &    6.6 &   5.6\hspace{10pt} & $-$\hspace{10pt} &  \\ 
 12 10 34.77  &  30 23 58.3 & NGC 4150 & 12.8 &  0.28 & 1638 &    107.7 &    9.0 &  10.8\hspace{10pt} & $-$\hspace{10pt} &  \\ 
 12 15 10.91  &  20 39 12.4 & NGC 4204 &  7.9 &  0.59 & 7092 &     40.3 &    5.4 &   1.4\hspace{10pt} & $-$\hspace{10pt} &  \\ 
 12 18 43.85  &  47 17 31.5 & NGC 4258 &  7.7 &  0.65 & 1618 &    403.3 &   17.6 &   1.2\hspace{10pt} & 1.6\hspace{10pt} &  \\ 
 12 18 57.84  &  47 16 07.1 & NGC 4258 &  7.7 &  0.34 & 1618 &    856.4 &   24.5 &   2.5\hspace{10pt} & 5.3\hspace{10pt} &  \\ 
 12 26 01.44  &  33 31 31.1 & NGC 4395 &  4.2 &  0.43 & 402 &    158.1 &   10.9 &   2.4\hspace{10pt} & 5.5\hspace{10pt} &  \\ 
\enddata
\end{deluxetable}
} 

\clearpage
{\tiny
\begin{deluxetable}{cccrrrrrrrc}
\tablecolumns{11}
\tablewidth{0pt}
\tablenum{1} 
\tablecaption{{\sl Continued}} 
\tablehead{
\colhead{R.A.} &
\colhead{Decl.} &
\colhead{Galaxy} &
\colhead{$D$} &
\colhead{$f_{\rm D25}^a$} &
\colhead{ObsID} &
\colhead{cts} &
\colhead{$S/N$} &
\colhead{$L_{{\rm cts}}^b/10^{39}$} &
\colhead{$L_{{\rm X}}^c/10^{39}$} &
\colhead{Ref.$^d$} \\
\colhead{(h m s)} &
\colhead{(d m s)} &
\colhead{} &
\colhead{(Mpc)} &
\colhead{} &
\colhead{} &
\colhead{} & \colhead{}
& \colhead{(erg s$^{-1}$)} &
\colhead{(erg s$^{-1}$)} &
\colhead{}
}
\startdata
 12 28 17.83  &  44 06 33.9 & NGC 4449 &  3.7 &  0.45 & 2031 &   1439.1 &   32.8 &   0.8\hspace{10pt} & 1.5\hspace{10pt} &  \\ 
 12 30 29.55  &  41 39 27.6 & NGC 4490 &  7.8 &  0.49 & 1579 &    276.5 &   14.4 &   0.9\hspace{10pt} & 19.8\hspace{10pt} &  \\ 
 12 30 30.56  &  41 41 42.3 & NGC 4485 &  7.8 &  0.35 & 1579 &   2232.7 &   32.4 &   7.3\hspace{10pt} & 6.3\hspace{10pt} &  \\ 
 12 30 30.82  &  41 39 11.5 & NGC 4490 &  7.8 &  0.39 & 1579 &    681.8 &   22.2 &   2.2\hspace{10pt} & 6.5\hspace{10pt} &  \\ 
 12 30 32.27  &  41 39 18.1 & NGC 4490 &  7.8 &  0.33 & 1579 &    509.4 &   19.2 &   1.7\hspace{10pt} & 2.9\hspace{10pt} &  \\ 
 12 30 36.32  &  41 38 37.8 & NGC 4490 &  7.8 &  0.01 & 1579 &    547.2 &   19.8 &   1.8\hspace{10pt} & 3.5\hspace{10pt} &  \\ 
 12 30 43.26  &  41 38 18.4 & NGC 4490 &  7.8 &  0.50 & 1579 &    979.3 &   26.7 &   3.2\hspace{10pt} & 8.3\hspace{10pt} &  \\ 
 12 32 42.83  &  00 06 53.2 & NGC 4517 &  9.8 &  0.18 & 0203170301 &   $-$ &  $-$ &   $-$\hspace{10pt} & 10.4\hspace{10pt} & 3 \\ 
 12 35 51.71  &  27 56 04.1 & NGC 4559 & 10.3 &  0.88 & 2027 &   2077.5 &   39.1 &  21.4\hspace{10pt} & 17.9\hspace{10pt} &  \\ 
 12 35 57.79  &  27 58 07.4 & NGC 4559 & 10.3 &  0.16 & 2027 &    256.9 &   14.2 &   2.6\hspace{10pt} & 3.9\hspace{10pt} &  \\ 
 12 35 58.56  &  27 57 41.9 & NGC 4559 & 10.3 &  0.10 & 2027 &   1392.2 &   33.0 &  14.4\hspace{10pt} & 16.8\hspace{10pt} &  \\ 
 12 36 08.90  &  19 19 55.9 & NGC 4561 & 12.3 &  0.82 & 7125 &     53.0 &    6.3 &   2.5\hspace{10pt} & $-$\hspace{10pt} &  \\ 
 12 41 29.14  &  41 07 57.7 & NGC 4618 &  7.3 &  0.62 & 7147 &    234.3 &   13.0 &   1.4\hspace{10pt} & 1.8\hspace{10pt} &  \\ 
 12 41 52.72  &  41 16 31.7 & NGC 4625 &  8.2 &  0.10 & 7098 &     26.2 &    4.4 &   1.1\hspace{10pt} & $-$\hspace{10pt} &  \\ 
 12 41 55.56  &  32 32 16.9 & NGC 4631 &  7.6 &  0.34 & 797 &   3164.0 &   47.9 &   3.2\hspace{10pt} & 3.6\hspace{10pt} &  \\ 
 12 42 11.13  &  32 32 35.9 & NGC 4631 &  7.6 &  0.10 & 797 &   1182.1 &   30.3 &   1.2\hspace{10pt} & 21.8\hspace{10pt} &  \\ 
 12 50 25.70  &  25 31 29.8 & NGC 4725 & 11.9 &  0.33 & 2976 &    217.3 &   12.5 &   1.3\hspace{10pt} & 1.2\hspace{10pt} &  \\ 
 12 50 26.37  &  25 33 19.4 & NGC 4725 & 11.9 &  0.71 & 2976 &    281.6 &   14.7 &   1.7\hspace{10pt} & 1.7\hspace{10pt} &  \\ 
 12 50 36.88  &  25 30 28.4 & NGC 4725 & 11.9 &  0.54 & 2976 &    239.5 &   13.6 &   1.4\hspace{10pt} & 1.2\hspace{10pt} &  \\ 
 12 50 52.72  &  41 07 19.0 & NGC 4736 &  5.2 &  0.02 & 808 &   2546.6 &   43.2 &   1.5\hspace{10pt} & 1.2\hspace{10pt} &  \\ 
 12 50 53.32  &  41 07 14.0 & NGC 4736 &  5.2 &  0.01 & 808 &   4523.7 &   56.8 &   2.7\hspace{10pt} & 3.1\hspace{10pt} &  \\ 
 12 55 12.31  &  00 07 51.9 & UGC 8041 & 14.2 &  0.58 & 7132 &     45.3 &    6.0 &   2.1\hspace{10pt} & $-$\hspace{10pt} &  \\ 
 13 15 19.54  &  42 03 02.3 & NGC 5055 &  9.2 &  0.90 & 2197 &   2648.6 &   43.9 &   8.3\hspace{10pt} & 30.4\hspace{10pt} &  \\ 
 13 15 39.33  &  42 01 53.4 & NGC 5055 &  9.2 &  0.31 & 2197 &    173.5 &   11.5 &   0.5\hspace{10pt} & 1.4\hspace{10pt} &  \\ 
 13 16 02.27  &  42 01 53.6 & NGC 5055 &  9.2 &  0.42 & 2197 &    648.5 &   22.1 &   2.0\hspace{10pt} & 12.2\hspace{10pt} &  \\ 
 13 29 38.61  &  58 25 05.6 & NGC 5204 &  4.5 &  0.18 & 2029 &   1467.5 &   32.3 &   3.4\hspace{10pt} & 3.3\hspace{10pt} &  \\ 
 13 29 43.30  &  47 11 34.8 & NGC 5194 &  8.4 &  0.46 & 1622 &    334.0 &   15.8 &   0.9\hspace{10pt} & 3.5\hspace{10pt} &  \\ 
 13 29 50.67  &  47 11 55.2 & NGC 5194 &  8.4 &  0.12 & 1622 &    242.8 &   13.3 &   0.7\hspace{10pt} & 6.9\hspace{10pt} &  \\ 
 13 29 53.32  &  47 10 42.7 & NGC 5194 &  8.4 &  0.21 & 1622 &    622.6 &   21.2 &   1.7\hspace{10pt} & 2.2\hspace{10pt} &  \\ 
 13 29 53.72  &  47 14 35.7 & NGC 5195 &  8.4 &  0.67 & 1622 &    421.4 &   18.0 &   1.2\hspace{10pt} & 2.1\hspace{10pt} &  \\ 
 13 29 57.57  &  47 10 48.6 & NGC 5194 &  8.4 &  0.24 & 1622 &    300.4 &   15.0 &   0.8\hspace{10pt} & 1.3\hspace{10pt} &  \\ 
 13 30 01.02  &  47 13 43.9 & NGC 5194 &  8.4 &  0.59 & 3932 &    522.0 &   20.4 &   2.6\hspace{10pt} & 3.9\hspace{10pt} & \\ 
 13 30 05.99  &  47 15 42.3 & NGC 5195 &  8.4 &  0.41 & 1622 &    390.3 &   16.9 &   1.1\hspace{10pt} & 1.0\hspace{10pt} &  \\ 
 13 30 07.54  &  47 11 06.1 & NGC 5194 &  8.4 &  0.69 & 1622 &    937.8 &   26.9 &   2.6\hspace{10pt} & 3.1\hspace{10pt} &  \\ 
 14 03 32.39  &  54 21 02.9 & NGC 5457 &  7.0 &  0.21 & 934 &   9003.5 &   73.3 &   4.7\hspace{10pt} & 14.4\hspace{10pt} &  \\ 
 14 04 14.29  &  54 26 03.8 & NGC 5457 &  7.0 &  0.76 & 934 &   4606.2 &   59.5 &   2.4\hspace{10pt} & 2.4\hspace{10pt} &  \\ 
 14 04 59.73  &  53 38 09.1 & NGC 5474 &  6.8 &  0.68 & 9546 &   5161.8 &   63.3 &   8.3\hspace{10pt} & 14.2\hspace{10pt} &  \\ 
 14 19 39.39  &  56 41 37.8 & NGC 5585 &  5.7 &  0.85 & 7150 &    243.9 &   14.0 &   1.6\hspace{10pt} & 2.5\hspace{10pt} &  \\ 
 15 15 58.60  &  56 18 10.0 & NGC 5907 & 13.4 &  0.29 & 0145190101 &   $-$ &  $-$ &   $-$\hspace{10pt} & 34.0\hspace{10pt} & 4 \\ 
 20 34 36.48  &  60 09 30.5 & NGC 6946 &  5.5 &  0.40 & 1043 &    941.7 &   26.6 &   0.7\hspace{10pt} & 3.1\hspace{10pt} &  \\ 
 20 34 52.32  &  60 09 11.7 & NGC 6946 &  5.5 &  0.01 & 1043 &    837.6 &   24.7 &   0.6\hspace{10pt} & 2.0\hspace{10pt} &  \\ 
 20 34 56.49  &  60 08 34.1 & NGC 6946 &  5.5 &  0.16 & 1043 &    406.3 &   17.8 &   0.3\hspace{10pt} & 3.3\hspace{10pt} &  \\ 
 20 35 00.74  &  60 11 30.8 & NGC 6946 &  5.5 &  0.45 & 1043 &   9034.0 &   82.4 &   6.8\hspace{10pt} & 5.7\hspace{10pt} &  \\ 
 20 35 18.79  &  60 10 56.6 & NGC 6946 &  5.5 &  0.73 & 1043 &    502.4 &   18.6 &   0.4\hspace{10pt} & 4.5\hspace{10pt} &  \\ 
 21 03 30.08  &  29 55 15.5 & NGC 7013 & 14.2 &  0.83 & 7130 &     18.4 &    3.7 &   1.1\hspace{10pt} & $-$\hspace{10pt} &  \\ 
 21 03 33.93  &  29 53 17.8 & NGC 7013 & 14.2 &  0.34 & 7130 &    104.6 &    9.0 &   6.5\hspace{10pt} & $-$\hspace{10pt} &  \\ 
 21 03 35.49  &  29 52 26.4 & NGC 7013 & 14.2 &  0.78 & 7130 &     23.8 &    4.4 &   1.5\hspace{10pt} & $-$\hspace{10pt} &  \\ 
 22 37 05.64  &  34 26 53.5 & NGC 7331 & 14.5 &  0.49 & 2198 &    122.7 &    9.5 &   1.0\hspace{10pt} & 3.2\hspace{10pt} &  \\ 
 22 37 06.61  &  34 26 20.1 & NGC 7331 & 14.5 &  0.47 & 2198 &    174.6 &   11.6 &   1.5\hspace{10pt} & 3.7\hspace{10pt} &  \\ 
 22 37 08.08  &  34 26 00.3 & NGC 7331 & 14.5 &  0.56 & 2198 &    201.1 &   12.2 &   1.7\hspace{10pt} & 2.4\hspace{10pt} &  \\ 
\enddata

\tablenotetext{a}{Radial position within host galaxy expressed as fraction of $D_{25}$ radius.}
\tablenotetext{b}{Unabsorbed 0.3$-$10.0~keV luminosity estimated from source counts
detected in the 0.3$-$6.0~keV band and assuming an absorbed power law spectrum of $\Gamma=1.7$ and a Galactic column density}
\tablenotetext{c}{Unabsorbed 0.3$-$10.0~keV luminosity estimated from model fits to the X-ray spectral energy distribution as give by the references in column 11}
\tablenotetext{d}{{\sc References} -- (1) Feng \& Kaaret (2009); (2) Roberts \etal\ (2004);
(3) Walton \etal\ (2011a), (4) Sutton \etal\ (2011)}

\end{deluxetable}
} 

\clearpage

{\small
\begin{center}
\vspace{12pt}
Table 2. ULX Luminosity Function Fit Parameters

\begin{tabular}{cccccccc} 
\hline \hline
 & \multicolumn{4}{c}{$C L_{\rm X}^{-\alpha} \exp(-L_{\rm X}/L_{c})$}  &
\multicolumn{3}{c}{$C L_{\rm X}^{-\alpha}$}\\
 & $C$ & $\alpha$ & $L_c^a$ & $\ell^b$ &  $C$ & $\alpha$& $\ell$ \\
 \hline
$N(>L_{sp})$  &  91.9 & 0.5 & 20.1 & 9.13\hspace{20pt} &  92.6 & 0.8 & 34.61\\
$N(>L_{cts})$ & 103.3 & 0.9 & 12.9 & 5.68\hspace{20pt}  & 92.9 & 1.1  & 14.90\\
$dN/dL_{sp}$ & 19.6$^{+13.7}_{-7.8}$ & 0.8$\pm$0.2 & 17.3$^{+18.7}_{-6.1}$ & 5.22\hspace{20pt}  & 32.3$^{+12.4}_{-9.5}$ & 1.4$\pm$0.2 & 16.41\\
$dN/dL_{cts}$ & 78.0$^{+124.5}_{-46.7}$ & 1.6$\pm$0.3 & 15.2$^{+53.0}_{-8.6}$ & 0.15\hspace{20pt}  & 92.0$\pm$25.0 & 2.0$\pm$0.2 & 3.61\\
\hline
\multicolumn{8}{l}{{\sc Notes}~---~$^{\rm a}$ all luminosities are in units of 10$^{39}$~\ergl; $^{\rm b}$ Maximum likelihood statistic}\\
\end{tabular}
\end{center}
} 

\end{document}